\documentclass[11pt]{article}
\usepackage{amssymb}
\usepackage{amssymb}
\usepackage{amsmath}
\usepackage[all]{xy}
\usepackage[dvips]{graphicx}
\numberwithin{equation}{section}

\def\be{\begin{equation}}
\def\ee{\end{equation}}
\def\ba{\begin{align}}
\def\ea{\end{align}}

\def\p{\partial}

\def\yboxit#1#2{\vbox{\hrule height #1 \hbox{\vrule width #1
\vbox{#2}\vrule width #1 }\hrule height #1 }}
\def\fillbox#1{\hbox to #1{\vbox to #1{\vfil}\hfil}}
\def\ybox{{\lower 1.3pt \yboxit{0.4pt}{\fillbox{8pt}}\hskip-0.2pt}}
\def\comments#1{}

\def\p{\partial}

\def\half{\frac{1}{ 2}}

\def\diag{{\rm diag}}

\def\cH{{\cal H}}

\def\CN{{\cal N}}

\def\II{\relax{I\kern-.10em I}}

\def\IZ{\relax\ifmmode\mathchoice
{\hbox{\cmss Z\kern-.4em Z}}{\hbox{\cmss Z\kern-.4em Z}}
{\lower.9pt\hbox{\cmsss Z\kern-.4em Z}}
{\lower1.2pt\hbox{\cmsss Z\kern-.4em Z}}\else{\cmss Z\kern-.4em
Z}\fi}
\def\IB{\relax{\rm I\kern-.18em B}}
\def\IC{{\relax\hbox{$\inbar\kern-.3em{\rm C}$}}}
\def\ID{\relax{\rm I\kern-.18em D}}
\def\IE{\relax{\rm I\kern-.18em E}}
\def\IF{\relax{\rm I\kern-.18em F}}
\def\IG{\relax\hbox{$\inbar\kern-.3em{\rm G}$}}
\def\IGa{\relax\hbox{${\rm I}\kern-.18em\Gamma$}}
\def\IH{\relax{\rm I\kern-.18em H}}
\def\II{\relax{\rm I\kern-.18em I}}
\def\IK{\relax{\rm I\kern-.18em K}}
\def\IN{\relax{\rm I\kern-.18em N}}
\def\IP{\relax{\rm I\kern-.18em P}}

\def\inbar{\,\vrule height1.5ex width.4pt depth0pt}

\def\p{\partial}

\font\cmss=cmss10 \font\cmsss=cmss10 at 7pt
\def\IR{\relax{\rm I\kern-.18em R}}

\def\CN{{\cal N}}

\def\lp10{l_P^{10}}
\def\lp11{l_P^{11}}
\def\R11{R_{11}}

\font\manual=manfnt
\def\dbend{\lower3.5pt\hbox{\manual\char127}}

\def\IZ{\relax\ifmmode\mathchoice
{\hbox{\cmss Z\kern-.4em Z}}{\hbox{\cmss Z\kern-.4em Z}}
{\lower.9pt\hbox{\cmsss Z\kern-.4em Z}} {\lower1.2pt\hbox{\cmsss
Z\kern-.4em Z}}\else{\cmss Z\kern-.4em Z}\fi}
\def\half {\frac{1}{ 2}}

\def\p{\partial}

\def\bar{\overline}

\def\CN{{\cal N}}

\def\rt2{\sqrt{2}}
\def\irt2{\frac{1}{\sqrt{2}}}

\def\b{\beta}

\font\cmss=cmss10
\font\cmsss=cmss10 at 7pt
\def\IL{\relax{\rm I\kern-.18em L}}
\def\IH{\relax{\rm I\kern-.18em H}}
\def\IR{\relax{\rm I\kern-.18em R}}
\def\inbar{\vrule height1.5ex width.4pt depth0pt}
\def\IC{\relax\hbox{$\inbar\kern-.3em{\rm C}$}}
\def\rlx{\relax\leavevmode}
\def\ZZ{\rlx\leavevmode\ifmmode\mathchoice{\hbox{\cmss Z\kern-.4em Z}}
 {\hbox{\cmss Z\kern-.4em Z}}{\lower.9pt\hbox{\cmsss Z\kern-.36em Z}}
 {\lower1.2pt\hbox{\cmsss Z\kern-.36em Z}}\else{\cmss Z\kern-.4em
 Z}\fi}
\def\IZ{\relax\ifmmode\mathchoice
{\hbox{\cmss Z\kern-.4em Z}}{\hbox{\cmss Z\kern-.4em Z}}
{\lower.9pt\hbox{\cmsss Z\kern-.4em Z}}
{\lower1.2pt\hbox{\cmsss Z\kern-.4em Z}}\else{\cmss Z\kern-.4em
Z}\fi}

\font\manual=manfnt
\def\dbend{\lower3.5pt\hbox{\manual\char127}}

\def\IZ{\relax\ifmmode\mathchoice
{\hbox{\cmss Z\kern-.4em Z}}{\hbox{\cmss Z\kern-.4em Z}}
{\lower.9pt\hbox{\cmsss Z\kern-.4em Z}} {\lower1.2pt\hbox{\cmsss
Z\kern-.4em Z}}\else{\cmss Z\kern-.4em Z}\fi}
\def\half {\frac{1}{ 2}}

\def\bar{\overline}

\def\rt2{\sqrt{2}}
\def\irt2{\frac{1}{\sqrt{2}}}

\title{The Topological Cigar Observables}

\author{Sujay K. Ashok$^{a}$ and Jan Troost$^{b}$ }
\date{}
\begin{document}
\maketitle

\begin{center}
$^{a}$Perimeter Institute for Theoretical Physics\\
Waterloo, Ontario, ON N$2$L$2$Y$5$, Canada \\
$^{b}$Laboratoire de Physique Th\'eorique\footnote{Unit\'e Mixte du CRNS et de l'Ecole
  Normale
Sup\'erieure, UMR 8549.}, Ecole Normale Sup\'erieure \\
$24$ Rue Lhomond Paris $75005$, France
\end{center}
\begin{abstract}
We study the topologically twisted cigar, namely the $SL(2,\IR)/U(1)$
 superconformal field theory at arbitrary level, and find the BRST cohomology of the topologically twisted $N=2$ theory. We find a one to one correspondence between the spectrum of the
twisted coset and singular vectors in the Wakimoto modules constructed over
the $SL(2,\IR)$ current algebra. The topological cigar cohomology is the
crucial ingredient in calculating the closed string spectrum of topological
strings on non-compact Gepner models. 
\end{abstract}

\section{Introduction}
String theory is able to describe dynamics on highly curved space-times. Using
exactly solvable conformal field theories, Gepner constructed models with $N=2$
supersymmetry in four dimensions, that describe string propagation on highly
curved compact backgrounds which are special points in the moduli space of
Calabi-Yau compactifications. 

To code the non-trivial physics one is interested in,
it often suffices to concentrate on a region near a
singularity, which may be embedded in a non-compact Calabi-Yau
manifold. It then becomes interesting to describe and study exact conformal field
theories that describe special points in the ``moduli space" of non-compact
Calabi-Yau's. 

Furthermore, if sufficient supersymmetry is preserved, the background allows us to study a BPS
sector of the full string theory, the topological string theory, which can be
defined for any flat $3+1$ dimensional background tensored with a central
charge $c=9$ and $N=2$ superconformal field theory. The observables of the
topological string theory lie in the cohomology of the BRST-operator of the
twisted theory (one of the worldsheet supercharges
 of the untwisted model). We focus in this
paper on the key ingredient to computing this cohomology, which is the
calculation of the cohomology of the relevant non-rational conformal field
theory.

A very large class of non-compact Gepner models (though not all \cite{Kiritsis:1993pb, Hori:2002cd}) is
obtained by tensoring $N=2$ minimal models with $N=2$ Liouville theories
(either in heterotic or type II string theory -- we concentrate on the latter
here). Equivalently \cite{Giveon:1999px, Hori:2001ax, Tong:2003ik, Israel:2004jt}, 
we can tensor $N=2$ minimal models with
superconformal cigar conformal field theories. These theories come equipped
with an $N=2$ superconformal algebra on the worldsheet and one can construct
four dimensional supersymmetric string backgrounds provided the internal CFT has a total central charge $c=9$. This constraint leaves us with a very large class of non-compact Gepner models (see
e.g. \cite{kutasov, Mizoguchi:2000kk, Eguchi:2000tc, Murthy:2003es,
  Eguchi:2004yi, Israel:2004ir, Eguchi:2004ik}).
 Note
that in general, the level $k$ of the cigar theory can 
take on integer, fractional, and even irrational values (when combined with yet another cigar theory at irrational central charge).

To be more concrete, we give a few simple examples that illustrate these general ideas.
Lower dimensional or non-critical superstring backgrounds in even dimensions
have been extensively studied since they were first discussed in
\cite{kutasov}. The conformal field theory (CFT) that describes these
non-singular non-critical string vacua are a product of the flat space theory in $d$ dimensions and an $\CN=2$ supersymmetric  generalization of the cigar background:  
\be\label{onefactor}
\IR^{1,d-1} \times \left[\frac{SL(2,\IR)}{U(1)} \right]_k \times \ldots 
\ee
In the minimal case in which there are no other factors present, the level of
the coset CFT is fixed to be a rational number by demanding absence of the
conformal anomaly at the quantum level, i.e. by requiring $c=15$. 
This fixes the supersymmetric level $k$ in terms of the dimension $d$ 
\be
\frac{3d}{2} + \left(3+\frac{6}{k}\right) = 15 \,. 
\ee
Moreover, one can also consider backgrounds in which the level is not uniquely fixed by
the constraint on the central charge. This simplest example is the class of vacua
\be\label{fourd}
\IR^{1,3} \times \left[\frac{SL(2,\IR)}{U(1)} \right]_{k} \times \left[\frac{SU(2)}{U(1)} \right]_{m}
\ee
has $c=15$ only if the supersymmetric
level $k$ of the cigar is given by the fraction
\be
k = \frac{2m}{m+2} \qquad m = 2,3,4 \ldots 
\ee
The level of the $SU(2)$-coset is quantized, as it descends from a compact group. 

These
backgrounds are especially interesting in the context of non-critical
holography \cite{Giveon:1999px, giveon2} in which these closed string
backgrounds 
are dual to certain non-gravitational theories obtained by taking a double
scaling limit of strings on non-compact Calabi-Yau manifolds of the form
\be
x^2+y^2+u^2+w^m = \mu \,.
\ee
The most well studied example is that of the deformed conifold ($m=2$ or
$k=1$). For this case, it has been proven \cite{Mukhi, Ghoshal} that the
B-model on the deformed conifold has a worldsheet description as a twisted
supercoset $SL(2,\IR)/U(1)$ at level $k=1$. A similar relationship for the
more general models in \eqref{fourd} has also recently been discussed in
\cite{bertoldi}, by relating these models
 to double scaling limits of certain matrix models.  

Good progress in the $k=1$ case was possible because the full set
of observables in the topologically twisted coset CFT was obtained by
E. Frenkel in the appendix to \cite{Mukhi}.  In this paper, we address in
detail the cohomology of the cigar at any level $k$ and exhibit the complete set of
topological observables by adapting the purely algebraic techniques used in 
E. Frenkel's appendix to \cite{Mukhi}. We will find that the topological observables are in one to one
correspondence with the singular vectors in the Wakimoto modules constructed
 over the $SL(2,\IR)$ current algebra. This is reminiscent of
 statements in earlier work on the relation between $N=2$
topological singular vectors, and their relation to $SL(2,\IR)$ singular vectors
\cite{Gato-Rivera:1992fd, Semikhatov, Feigin:1997ha}. 

These non compact Gepner models are also intimately connected with some
 bosonic string theories, the most well-known example being the relationship
 between the $k=1$ cigar and the $c=1$ string \cite{Mukhi}. More recently, a particular 
 class of correlation functions in the (integer) level $k$ cigar times a level
 $k$ minimal model has been mapped to those in the $(k,1)$ bosonic minimal
 string \cite{Takayanagi, Sahakyan, Niarchos} \footnote{See also \cite{Rastelli} which
 relates twisted AdS$_3 \times S^3$ with $k$ units of flux through the $S^3$
 to the bosonic $(k,1)$ minimal string.}. The results we obtain should be 
instrumental in further clarifying the relation between topological cosets and bosonic string theories. 

\subsubsection*{Organization}
In section \ref{introtocigar}, we review the worldsheet formulation of the
topologically twisted supercoset conformal field theory. 
In section \ref{closedcohom}, we compute the closed string
cohomology of the twisted supercoset. We first discuss the integer level case
in some detail in section \ref{integerk} to clarify the techniques we use. The
general proof for the rational level case is given in section
\ref{positiverational}.  An appendix discusses a concrete example in detail
which illustrates the abstract proof in the main text. It also contains some
of the background in affine algebra that is useful to fully prove the
statements in
 the main text.

\section{The twisted cigar at level $k$}
\label{introtocigar}

\subsection{The $\CN=2$ superconformal symmetry}\label{susycigar}
To begin with, we consider the $\CN=2$ supersymmetric topologically twisted
 cigar at level $k$. We briefly review the construction of the conformal field theory
and the twisting procedure.

The $\CN=1$ currents of the parent $SL(2,\IR)$ theory at level $k $ has affine
currents $J^a$ and fermions $\psi^a$ that satisfy the OPE
\begin{align}
J^a(z)\, J^b(w) &\sim \frac{g^{ab}\, k/2}{(z-w)^2}+\frac{f^{ab}_c\, J^c}{z-w} \cr
J^a(z)\, \psi^b (w) &\sim \frac{i f^{ab}_c\, \psi^c}{z-w} \cr
\psi^a(z)\,\psi^b(w) &\sim \frac{g^{ab}}{z-w}  \,.
\end{align}
Our conventions are left-right symmetric, with $g_{ab} = \diag(+,+,-)$ and
$f^{123} = \bar{f}^{123} = 1$. In order to define the $\CN=2$ currents, we
first define\footnote{We follow the conventions in \cite{Mukhi}. 
With these conventions, $j^3=J^3+ \psi^+\psi^-$.}  
\be
j^a = J^a - \hat{J}^a = J^a - \frac{i}{2}f^a_{bc}\,\psi^b\,\psi_c  \,.
\ee 
The currents $j^a$ commute with the free fermions $\psi^a$
and generate a bosonic $SL(2,\IR)$ at level $k+2$. The Hilbert space of the
original $\CN=1$ $SL(2,\IR)$ model factorizes into a purely bosonic
$SL(2,\IR)$ at level $k+2$ and three free fermions. We now implement the coset
following Kazama-Suzuki \cite{Kazama}. The currents of the $\CN=2$
superconformal algebra are:
\begin{align}\label{nequalstwo}
T &= T_{SL(2,\IR)} - T_{U(1)} \qquad G^{\pm} = \sqrt{\frac{2}{k}}\, \psi^{\pm}\,j^{\pm}  \cr
J^R &= -\frac{2}{k}\,j^3+\left(1+\frac{2}{k}\right)\,\psi^+\, \psi^-  = -\frac{2}{k}\, J^3+ \psi^+\psi^-
\end{align}
where we have defined $\psi^{\pm} = \frac{\psi^1\pm i\psi^2}{\sqrt{2}}$ and $j^{\pm} = \frac{j^1\pm ij^2}{\sqrt{2}}$ and 
\be
T_{U(1)} = -\frac{1}{k}\, J^3 J^3 + \frac{1}{2}\psi^3\p \psi^3 \,.
\ee
One can check that these currents generate an $\CN=2$ superconformal algebra with central charge $c=3+\frac{6}{k}$. 

\subsection{Gauging}\label{gauging}
The axial gauging of the coset is
 done by adding an extra boson $X$ and superpartner $\psi^X$, and restricting to the cohomology of  an
 additional
gauging BRST charge $Q_{U(1)}$, whose left-moving current is
\begin{align}
J_{BRST} &= C\,(J^3- i\sqrt{\frac{k}{2}}\, \p X) + \gamma'\,(\psi^3 - \psi^X)  \,.
\end{align}
Here, $(B,C)$ is a $(1,0)$ ghost system
associated with this BRST symmetry with central charge $c=-2$. 
The field $X$ indicates the boson which
will be identified with the angular direction of the cigar geometry \cite{Witten, Dijkgraaf}. The
$\b',\gamma'$ superghosts combine with the fermions $\psi^3$
and $\psi^X$ to form a Kugo-Ojima quartet that decouples
from the cohomology \cite{Figueroa-O'Farrill:1995pv}. We can therefore safely ignore it in what follows.

The gauging current is defined to be
\be\label{gaugecurrent}
J_g = J^3 - i\sqrt{\frac{k}{2}}\p X  
\ee
{From} the definition, it is easy to check that it is a null-current and has
non-singular operator product expansions with the energy momentum tensor $T$,
and $U(1)_R$ current $J^R$, and also with itself. The BRST operator $Q_{U(1)}$
of
the cigar theory is defined to be 
\be
Q_{U(1)} = \oint dz\  J_{BRST}.
\ee
The cohomology with respect to the gauging BRST operator is standard to
compute \cite{Karabali:1988au}\cite{Hwang:1993nc}.
 The
calculation essentially follows from the $Q_{U(1)}$ exactness of both the 
total scaling operator, as well as the total charge in the gauged direction,
and from the fact these operators are diagonalizable in the Hilbert space.
Those facts imply that non-trivial cohomology elements carry no 
oscillator excitations in the original gauged direction, nor in the
auxiliary direction $X$. Also, the gauging locks the charge in the
direction $X$ to the charge in the gauged direction of the original model.
(There is a trivial degeneracy in the zero-mode of the ghosts which we can
safely ignore.) The details of
this cohomological calculations are discussed in \cite{Hwang:1993nc}. The
net result,
as explained, is according to intuition: excitations in the gauged directions 
are removed from the theory.

\subsection{Wakimoto Representation of $SL(2,\IR)$}
In what follows, we will make good use of the Wakimoto free field representation of the
$SL(2,\IR)$ currents:
\begin{align}
\label{wakimotocurrents}
j^- &= \beta  \cr
j^3 &= \beta\,\gamma + \sqrt{\frac{k}{2}}\, \p\phi  \cr
j^+ &= \beta\, \gamma^2 + \sqrt{2k}\, \gamma\, \p\phi + (k+2)\, \p\gamma.
\end{align}
The energy momentum tensor in these variables is given by 
\be
T_{(SL(2,\IR)} = \beta\,\p \gamma - \half (\p\phi)^2 -\frac{1}{\sqrt{2k}}\, \p^2\phi -\half \psi^+\p\psi^- - \half\psi^- \p\psi^+ \,.
\ee
We collect the various fields and some of their properties in Table \ref{untwistedtable}.
\begin{table}
\begin{center}
\begin{tabular}{|c|c|c|c|c|}
\hline
 Field & $\Delta$ & Q & c \\
\hline\hline
X & $-$  & $0$ & $1$  \\
\hline
$\phi$ & $-$ & $\sqrt{\frac{2}{k}}$ & $1+\frac{6}{k}$ \\
\hline
$(\psi^+,\psi^-)$ & $(1/2,1/2)$ & $-$ & $1$ \\
\hline
$(\b,\gamma)$ &$(1,0)$ & $-$ & $2$ \\
\hline
$(B,C)$ & $(1,0)$ & $-$ & $-2$ \\
\hline
 \end{tabular}
\end{center}
\caption{List of fields in the untwisted theory, their conformal weight $\Delta$, the background charge $Q$ for the bosons and the central charge $c$.\label{untwistedtable}}
\end{table}
\subsection{Twisting}\label{twisting}

We consider the topologically twisted theory whose stress tensor is defined by 
\be\label{ktwist}
T \longrightarrow T + \half \p J^R -\left(1-\frac{1}{k}\right) \p J_g 
\ee
where $J^R$ is the $R$-current that appears in \eqref{nequalstwo}. We have modified the usual definition of the twisted theory by adding a certain multiple times the gauging current $J_g$. As far as the physical obervables are concerned, this addition makes no difference, as all physical quantities are uncharged under the gauging current. However, the twist as defined in \eqref{ktwist} makes contact with early attempts  \cite{Ohta, Takayanagi} to relate the twisted coset theory to a bosonic $c < 1$ string theory (i.e. $c < 1$ matter coupled to $bc$ (reparametrization) ghosts). Let us see how this comes about : using the explicit expressions for the currents in terms of the Wakimoto free fields, the twisted energy momentum tensor in \eqref{ktwist} can be written as
\begin{align}\label{twistedT}
T = - \p \b \gamma - \half (\p\phi)^2 - \frac{k+1}{\sqrt{2k}}\, &\p^2\phi - \half(\p X)^2 +i \frac{k-1}{\sqrt{2k}}\, \p^2 X  \cr
&- \half\psi^+\p\psi^- -\half \psi^-\p\psi^+  +\frac{3}{2}\p(\psi^+\psi^-) 
\end{align}
We collect in Table \ref{twistedtable}, the conformal dimensions and central charges of the twisted theory. We remark here that the precise combination of $R$-current and gauging current in \eqref{ktwist} was chosen to get the coefficient of the last term in \eqref{twistedT} to be precisely $3/2$. From the table, we see that this has caused the fermions on the cigar $(\psi^+,\psi^-)$ to have spins $(-1,2)$.  
\begin{table}[h]
\begin{center}
\begin{tabular}{|c|c|c|c|}
\hline
 Field & $\Delta$ & Q & c \\
\hline\hline
X & $-$  & $-i\left(\sqrt{2k}-\sqrt{\frac{2}{k}}\right)$ & $1-6\frac{(k-1)^2}{k}$  \\
\hline
$\phi$ & $-$ & $\sqrt{\frac{2}{k}} + \sqrt{2k} $ & $1+6\frac{(k+1)^2}{k}$ \\
\hline
$(\psi^+,\psi^-)$ & $(-1,2)$ & $-$ & $ - 26$ \\
\hline
$(\b,\gamma)$ &$(0,1)$ & $-$ & $2$ \\
\hline
$(B,C)$ & $(1,0)$ & $-$ & $-2$ \\
\hline
 \end{tabular}
\end{center}
\caption{List of fields in the twisted theory, their conformal weight $\Delta$, the background charge $Q$ for the bosons and the central charge $c$.\label{twistedtable}}
\end{table}
Furthermore, we observe that the field content of the twisted theory is identical to the free field formulation of the bosonic $c < 1$ string theory by identifying $X$ with the matter field, $\phi$ with the Liouville field \cite{Bouwknegt}, and the fermions on the cigar with the $(b,c)$ ghosts.

The above remark indicates that
 the results we obtain are pertinent to the tentative correspondence
 with bosonic string theories that was mentioned in the introduction (but we leave
 this question for future work). We continue to work with 
the $(\b, \gamma, \phi, X, b, c)$ variables in what follows, and
will always consider states in the Hilbert space
\be\label{hilbert}
\cH_{sl_2} \otimes \cH_{b,c}\otimes \cH_{X}
\ee
where $sl_2$ in the subscript refers to the free field (Wakimoto) module of
 the bosonic $SL(2,\IR)$ algebra at level $k+2$ in terms of the $(\b, \gamma,
 \phi)$ variables, and $\cH_{bc}$ and $\cH_{X}$ are the usual Fock spaces of
 the fermions and bosons. As we will see, an important grading on states in
 this product Hilbert space will be provided by 
the fermion number (charge under the $bc$ current) which we will henceforth
 refer to as the ghost number, in analogy with the bosonic string.

\subsection{The total BRST operator}
In the topologically twisted cigar theory, the BRST currents are given by
\be\label{topBRST}
Q_{top} := Q^+ = \oint G^+ = \oint \psi^+ j^+ = \oint c\, j^+ 
\ee
while the other twisted supercurrents are identified with
\be\label{bghosts}
G^- = \psi^-\, j^- = b\,\beta 
\ee
where we have used the Wakimoto representation for the currents, and the renaming:
\be
j^{-} = \b\qquad \hbox{and}\qquad (\psi^{+},\psi^-)= (c,b)\,.
\ee
The quartet of the ghosts $(\b,\gamma)$ and $(B,C)$ have total central charge zero. 

We now restrict to the cohomology of the two BRST charges associated with the gauging and twisting, and define the operator
\be\label{Q}
Q = Q_{top} + Q_{U(1))} = \int dz \left[C\left(j^3-i\sqrt{\frac{k}{2}} \p X - cb\right) + cj^+ \right]  
\ee 
Thus, we have a sum of two commuting BRST operators. We already gave an
intuition for the role of the BRST term associated to gauging. In the
following we study the additional effect of the twisting on the set of observables,
 i.e. we compute the total BRST cohomology.

\section{Closed string cohomology}\label{closedcohom}
Our primary goal in this paper will be to compute the cohomology of Q in the
 Wakimoto modules $W_j$ for all positive values of the supersymmetric level $k$. The
 discussion is very much dependent on the submodule structure of the Wakimoto
 modules. In the following subsections, for the readers convenience, we will
 discuss cases with increasing
 difficulty, culminating in the general proof by induction for the most
 intricate case of rational level.

 We first collect a few general results about Wakimoto modules and
 their cohomology. As the explicit Wakimoto representation of
 the current algebra generators suggest (see equation
 (\ref{wakimotocurrents})), 
the Wakimoto modules $W_j$ are
 defined by embeddings of the $SL(2,\IR)$ current algebra into free fields
 such that the modules $W_j$ are free over the Lie subalgebra spanned by
 $j_n^3$ for $n < 0$ and $j^-_{n}$ for $n \le 0$ and co-free over
 the Lie subalgebra spanned by $j^-_n$ for $n >0$. (Co-free means that
 the dual representation is free with respect to this generator. Here, it
 is an invariant way of encoding that the operators $\gamma_{-n}$ act freely as
 creation operators on the vacuum.)   
\begin{itemize}
\item For $j \ge -\half$, the Wakimoto modules are isomorphic to the Verma modules built on the same highest weight state while for $j \le -\half$, it is the dual $W^*_j$ that is isomorphic to the 
Verma module. This was proven in \cite{Frenkel:1992ex} by explicitly comparing 
determinant formulae computed in the respective modules, and by constructing a
bijective map. (We use here that the supersymmetric level $k$ of the cigar is positive.)
\item It should be noted then that since we can map Wakimoto
  modules to Verma modules, and vice versa, the results that we will derive
  for Wakimoto modules in the following can straightforwardly be translated
  into complete results for the cohomology in Verma modules as well.
\item For the dual Wakimoto module, a lemma in the appendix of \cite{Mukhi} states that 
\be\label{dualwaki}
H^q(W^*_j) = \delta_{q,1} \{j\} 
\ee 
where $q$ refers to the ghost number (of the corresponding operator)
 mentioned earlier. The proof of this fact stated in \cite{Mukhi} can be found
 by 
noting that the scaling operators in the $(b,c)$ sector, as well as in the
$(\beta, \gamma)$ sectors are BRST exact. Since both operators are
diagonalizable, it follows that no non-trivial excitations in these directions
are allowed for cohomologically non-trivial operators. Also, no $\phi$
oscillators are then allowed, because of the gauging of the $U(1)$
current. Then, in this zero-level subspace, one can explicitly find the form
of the $Q_{top}$ operator, which reduces to its zero-mode term ($Q_{top}=c_0
j^+_0$). It is then easy to show by direct calculation that the cohomology
does indeed localize as claimed in the lemma in the appendix of \cite{Mukhi},
for any level $k$, and any dual  Wakimoto module (see equation 
(\ref{dualwaki})).\footnote{Explicitly, the one non-trivial
state in the cohomology is the highest weight state of the dual Wakimoto module
tensored with the down ghost vacuum.}
\end{itemize} 
It turns out that the above rather straightforward calculation of a cohomology
in a dual Wakimoto module, in a specific free field realization, is the only
calculation that needs to be done explicitly. Indeed, we have now seen that
Wakimoto modules and Verma modules are interchangeable (in the
sense of \cite{Frenkel:1992ex}). 
Moreover, the embedding diagrams between Verma modules have been
constructed in the ground breaking work \cite{Malikov} (see also
\cite{Feigin:1990ut}). It will turn out that the short exact sequences
that follow from the embedding diagrams of Verma modules, along with the specific
cohomology computed above, contain sufficient information to compute the
cohomology for all Wakimoto and Verma modules.

\subsection{Singular vectors in Verma modules}
The technique we use to compute the closed string cohomology relies on knowing
 the submodule structure of
 a given Wakimoto module, and hence on the structure of embeddings of singular
 vectors in the module. 
Since Wakimoto modules are isomorphic to Verma modules or their duals, we can
 use the results on embedding diagrams for the latter. Recall that Verma
 modules can occur as submodules of a Verma module when they are built on
 singular vectors (which by definition are annihilated by all creation
 operators, and therefore constitute a new highest weight state). 
A  most useful result is then the location of the singular vectors within a
given Verma module. The Kac-Kazhdan
 formula \cite{Kac} states that a Verma module $V_j$ of spin $j$ over the (supersymmetric)
level  $k$ $SL(2,\IR)$ current algebra has singular vectors whenever the spin
 $j$ satisfies,
for some integer $r,s$:  
\be\label{kackazhdan}
2j+1 = r+ks \quad\hbox{such that}\quad rs > 0 \quad\hbox{or}\quad r >0, \, s=0 \,
 .
\ee
For each solution, there is a singular vector labeled by the spin $j'$ 
\begin{align}
j' = j -r \qquad \hbox{and} 
\end{align}
and at relative conformal dimension $\Delta h = rs$
with respect
 to the highest weight state
 in $V_j$. 
It will turn out (see subsection \ref{linear}) that when a
 Verma module has only one singular vector, the calculation of the cohomology
 is rather straightforward. It is also clear that in the case when there are at least two solutions
 to the Kac-Kazhdan equation for a given value of the spin $j$ and the level
 $k$, the level is rational. 
Thus, the case of irrational levels will be (implicitly) included when we
discuss
the general calculation for the case of a single singular vector.


\subsection{Linear submodule structure} 
\label{linear}
We start out our discussion with a very much restricted subset of all cases we
consider, in order to connect with results already obtained in the literature
first. Some steps of the derivation will be postponed until the full treatment
in the next subsections, in order to first motivate the use of the analysis.

For $j$ values that satisfy $2j+1\in k\IZ$, the submodule structure of the
Wakimoto modules is very similar to those for the level $k=1$, discussed in
\cite{Mukhi} and the singular vectors appear in a 
nested fashion. For $j\ge 0$, the embedding diagram for the Wakimoto module
$W_j=V_j$ is given by:
\be
V_{j} \longrightarrow V_{-j-1}  \longrightarrow V_{j-k} \longrightarrow V_{k-j-1} \ldots 
 \longrightarrow V_{-1/2}
 \ee
where we have shown the end point $V_{-1/2}$ of the embedding diagram 
that always arises for $j$ half-integer.
 For $j$ integer, the end point is always $j=-1$. 
We will derive this embedding diagram in more detail in the next sections.
Similarly, for $j < 0$, the sub-module structure of the Wakimoto module $W_j$ is as follows :
\be
V_j \longleftarrow V_{-j-k-1} \longleftarrow V_{j+k} \longleftarrow \ldots \longleftarrow V_{-1/2}
\ee
The analysis of the cohomology
is exactly as in the appendix to \cite{Mukhi} and we obtain the following result 
for the cohomology. (We restrict ourselves to the only non-trivial
$j$-values, which are either half-integer or integer.) We get:
\begin{align}\label{theorem}
\hbox{For}\ j= -\frac{k}{2},\ldots, -1, -\half: \quad H^1(W_j) &= \{j\} \cr
\hbox{For}\ j > -\half : \quad H^1(W_j) &= \{j,j-k, \ldots, k-j-1, -j-1\} \quad\hbox{and}\cr
H^2(W_j) &= \{j-k,j-2k \ldots k-j-1,-j-1\}  \cr
\hbox{For}\ j < -\frac{k}{2} : \quad H^1(W_j) &= \{-j-k-1 \ldots , j\} \quad\hbox{and}  \cr
H^0(W_j) &= \{-j-k-1, \ldots , j+k\} 
\end{align}
Following the discussion in \cite{Ashok:2005xc}, we can infer from these
spin values and the ghost number at which the cohomology arises, 
the values of the momentum of the auxiliary boson $X$,
 and write explicit cohomology representatives. It is easy to check that these
 coincide in the case at hand
 with the representatives constructed in \cite{Takayanagi}. It was also
 observed there that those representatives are valid only for $2j+1 \in
 k\IZ$, i.e. the case to which we restricted above. In the coming sections, we 
will see that these are only a subset of all possible cohomology elements, and
we now turn to finding these elements. Moreover, we give many more details of
the derivation of the cohomology, and the above case will be included as a
special case.

\subsection{General Result for integer level}\label{integerk}

In this section, we give a proof by induction for the cohomology of the BRST
operator
$Q$
in the Wakimoto modules for integer level $k$ generalizing the examples worked out in the
appendix (which can be consulted first by the reader prefering a warm-up). 
As we saw in the examples in the appendix, the submodule structure of $W_j$ takes the form 
\begin{equation}
\xymatrix{
& \bullet\ar[ddr]\ar[r] &\bullet &\bullet\ar[dr] & \\
V_j \ar[ur]\ar[dr]& &  \ldots &\ldots& V_{j_0} \\
&\bullet\ar[uur]\ar[r] &\bullet &\bullet\ar[ur] & 
}
\end{equation}
Here, the dots in the diagram indicate singular vectors that appear in either
the original module $V_j$, or those appearing in the submodules. Let us
elaborate a bit more on how the spin $j$-values of the dots can be derived. For $k$
integer, there is a simple way to generate all
spins $j$ appearing in the embedding at one go by using affine Weyl
reflections about the spin values $j=-\half$ and $j=-\frac{k+1}{2}$. 
We generate the diagram above using affine
Weyl reflections and show, by analyzing the Kac-Kazhdan equation, that the
$j$-values generated indeed correspond to all 
singular vectors.

By using reflections around the points $j=-\half$ and $j=-\frac{k+1}{2}$, we
can reflect any half-integer $j$ spin into the range
\be
-\frac{(k+1)}{2} \le j_0\le -\half \,. 
\ee
where $j_0$ denotes all possible end-points to an embedding diagram.
First of all, we remark that all $j_0$ within this bound have a trivial embedding diagram, consisting of a
single point (since there are no singular vectors as is easily checked from
the Kac-Kazhdan equation). The embedding diagrams for other $j$-values can be built up
inductively by adding a pair of dots at each step. The $j$-values of the successive dots
are obtained by affine Weyl reflections about the values $j=-\half$ and
$j=-\frac{(k+1)}{2}$. At the first level, this leads to the $j$ values
$\{-j_0-1, -j_0-k-1\}$. Affine Weyl reflections on these dots lead to the pair
$\{j_0-k, j_0+k\}$, $\{-j_0-1+k, -j_0-2k-1\}$ and so on. Thus, reading the
diagram backwards, the pairs follow the pattern $\{-j_0-1+nk,-j_0-1-(n+1)k\}$
followed by the pair $\{j_0-nk,j_0+nk\}$ and so on recursively. In this way,
embedding diagrams for all half-integer $j$ can be constructed, and the
precise labeling of the dots obtained. We exhibit the embedding diagram for
$j=-j_0-1+nk$ as an example. 
\begin{equation}\label{embedex}
\xymatrix{
& V_{j_0+nk}\ar[ddr]\ar[r]& V_{-j_0-1+(n-1)k}& V_{j_0+k}\ar[ddr]\ar[r]  & V_{-j_0-1}\ar[dr] & \\
V_{j} \ar[ur]\ar[dr]&& \ldots&   \ldots& & V_{j_0} \\
& V_{j_0-nk}\ar[uur]\ar[r]& V_{-j_0-1-nk}& V_{j_0-k}\ar[uur]\ar[r] &
V_{-j_0-k-1}\ar[ur] & 
\\
}
\end{equation}
We note that the case $2j+1 \in kZ$ is a special case of the above, in which
the double string embedding diagram collapses onto a single string embedding
diagram. That special case was treated in subsection \ref{linear}.

\subsubsection*{Singular vectors}
{From} the embedding diagram, it is clear that there are four cases of $j$-values to be considered (for $n > 0$) 
\begin{itemize}
\item $ j_0+nk$ : As mentioned earlier, singular vectors are solutions to the
  Kac-Kazhdan equation \eqref{kackazhdan}, i.e. pairs of integers $(r,s)$ such
  that $r+ks = 2j_0+2nk+1$. Let us discuss the solution set in some detail:
  taking $(r,s)=(2j_0+2nk+1,0)$, we get the singular vector labeled by $j=
  -j_0-1-nk$. Increasing $s$ by one, we get the solution $(r,s)=
  (2j_0+1+(2n-1)k,1)$ leading to the singular vector $j=
  -j_0-1-(n-1)k$. Proceeding this way, we find all $s$ values from $1$ up to
  $s=2n-1$, in which case we obtain the singular vector corresponding to
  $j=-j_0-1+(n-1)k$. If we define $S^+_{x,y}= \{j_0+xk, j_0+(x+1)k, \ldots ,
  j_0+yk\}$, and $S^-_{x,y} = \{-j_0-1+xk, -j_0-1+(x+1)k, \ldots ,
  -j_0-1+yk\}$, then the singular vectors corresponds to the list $S^-_{-n, n-1}$. 
\item $j_0-nk$ : A similar analysis leads to the set of singular vectors
  identical to the case above, and we get the list of singular vectors 
  $S^-_{-n,n-1}$.
\item $ -j_0-1+nk$ : We now get singular vectors of the form $j_0\pm mk$ with
 $0 \le m \le n$, which corresponds to the list $S^+_{-n,n}$. 
\item $ -j_0-1-nk$ : We get the list of singular vectors
 $S^+_{-n,n}$. We have thus verified that the embedding diagram of $V_{-j_0-1+nk}$ indeed coincides with \eqref{embedex}. 
\end{itemize}

\subsubsection*{Proof by induction}
We now  claim that the cohomology of the Wakimoto module is given in terms of the
 singular vectors appearing in the module. We will give the details of 
the proof
 in the case $j=-j_0-1+nk$. The other cases can be proven similarly. We
 will prove this claim by induction, by assuming that the result holds for all
 $m \le n-1$ and then show that the result also 
holds for $m=n$. For $j=-j_0-1+nk$, the embedding diagram is shown
 in \eqref{embedex}. In this case, the following short exact sequences are valid:
\begin{multline}\label{firstImsequence}
0\longrightarrow \hbox{Im}(V_{-j_0-1+(n-1)k}\oplus V_{-j_0-1-nk}) 
\longrightarrow V_{j_0+nk}\oplus V_{j_0-nk} \longrightarrow \hbox{Im}(V_{j_0+nk}\oplus V_{j_0-nk})\longrightarrow 0
\end{multline}
\begin{equation}\label{secondImsequence}
0\longrightarrow \hbox{Im}(V_{j_0+nk}\oplus V_{j_0-nk})\longrightarrow V_{-j_0-1+nk}\longrightarrow I_{-j_0-1+nk}\longrightarrow 0
\end{equation}  
where $I_j$ refers to the irreducible module obtained by modding out the
singular vectors in $V_j$. The second sequence indicates that when we mod out by
the leading singular vectors, the Verma module becomes irreducible, while the
first sequence quantifies the fact that in that modding, we have double counted
the submodules common to the two leading singular vectors.
Now, we observe that in order to compute the cohomology using induction, we also need to
postulate the cohomology of the various images of direct sums of Verma modules
in other Verma modules. We postulate the following ansatz for the Wakimoto modules
\begin{align}\label{ansatz}
H^1(W_{-j_0-1\pm mk}) &= S^+_{-m,m} \cup \{-j_0-1\pm mk\} \cr
H^2(W_{-j_0-1+mk}) = H^0(W_{-j_0-1-mk}) &=S^+_{-m,m} \cr
H^1(W_{j_0\pm mk}) &= S^-_{-m,m-1} \cup \{j_0\pm mk\} \cr 
H^2(W_{j_0+mk}) = H^0(W_{j_0-mk}) &=S^-_{-m,m-1}
\end{align}
and introduce the following ansatz for the cohomology of the image modules 
\begin{align}
H^1(\hbox{Im}(V_{-j_0-1+(n-1)k} \oplus V_{-j_0-1-nk})) &= \hbox{singular vectors in $V_{j_0\pm nk}$}\cr
& = S^-_{-n,n-1} \label{imageone}\\
H^2(\hbox{Im}(V_{-j_0-1+(n-1)k} \oplus V_{-j_0-1-nk})) &= \hbox{singular vectors in $V_{-j_0-1+(n-1)k}$} \cr
&= S^+_{-(n-1),(n-1)}\label{imagetwo}
\end{align}
where in the right hand side of
\eqref{imagetwo}, we could have used $V_{-j_0-1-nk}$ as well. Our ansatz can
be explained as follows: the ghost number one cohomology $H^1$ of the image of the sum of two Verma modules
$V_{j_1}\oplus V_{j_2}$ is given by the set of singular vectors in the Verma
module $V_{j}$ in which $j_1$ and $j_2$ are the leading singular
vectors. Similarly, cohomology at ghost number two $H^2$ of the image is given by the singular vectors in
either $V_{j_1}$ or $V_{j_2}$, which coincide in all cases we consider. We
will see that this ansatz generalizes 
to the rational level case also. 

We now have all the ingredients to complete our proof by induction. Let us
consider the short exact sequence in \eqref{firstImsequence}. By the induction
hypothesis, we know the cohomology of the first and middle element in the
sequence. Considering the long exact sequence of cohomology, and using the
induction hypothesis, the long exact 
sequence collapses to
\begin{multline}
0\longrightarrow S^-_{-n,n-1}\longrightarrow S^-_{-n,n-1}\,\cup\, \{j_0-nk, j_0+nk\}   
\longrightarrow H^1(\hbox{Im}(V_{j_0+nk}\oplus V_{j_0-nk})) \longrightarrow \cr
\longrightarrow S^+_{-(n-1),n-1} \longrightarrow S^-_{-n,n-1} \longrightarrow H^2(\hbox{Im}(V_{j_0+nk}\oplus V_{j_0-nk})) \longrightarrow 0
\end{multline}
Solving for the cohomology, we get $H^0(\hbox{Im}(V_{j_0+nk}\oplus V_{j_0-nk})) = 0 $,
\begin{align}
H^1(\hbox{Im}(V_{j_0+nk}\oplus V_{j_0-nk})) &= S^+_{-n,n} \cr
&= \hbox{singular vectors in $V_{-j_0-1+nk}$}  \label{interimageone}\\ 
H^2(\hbox{Im}(V_{j_0+nk}\oplus V_{j_0-nk})) &= S^-_{-n,n-1}\cr
&= \hbox{singular vectors in $V_{j_0\pm nk}$ }\label{interimagetwo}
\end{align}
where in \eqref{interimageone} and \eqref{interimagetwo}, we indicate that the
 result is what one would expect, given our ansatz for the cohomology of the
 image modules
 in \eqref{imageone} and \eqref{imagetwo}. 
We also find that the cohomology of the dual of the image is exactly the
 same as this one, with the exchange of ghost number zero $H^0$ and ghost
 number two $H^2$ cohomology. Let us now
 consider the short
 exact sequence dual to
\eqref{secondImsequence}: 
\be\label{dualseq}
0 \longrightarrow I_{-j_0-1+nk} \longrightarrow V^*_{-j_0-1+nk} \longrightarrow \hbox{Im}^*(V_{j_0+nk}\oplus V_{j_0-nk}) \rightarrow 0
\ee 
Note that although we have taken the dual sequence, the irreducible module has
remained unchanged, since it is canonically isomorphic to its dual.
The long exact sequence corresponding to \eqref{dualseq} 
now takes the form 
\begin{multline}
0 \longrightarrow H^0(I_{-j_0-1+nk}) \longrightarrow 0 \longrightarrow S^-_{-n,n-1} 
\longrightarrow H^1(I_{-j_0-1+nk})\longrightarrow \\
\longrightarrow\{-j_0-1+nk\} \longrightarrow S^+_{-n,n} \longrightarrow H^2(I_{-j_0-1+nk}) \longrightarrow 0
\end{multline}
which leads to 
\begin{align}
H^1(I_{-j_0-1+nk}) &= S^-_{-n,n} \cr
H^2(I_{-j_0-1+nk}) &= S^+_{-n,n} \,.
\end{align}
We now input all these results into the long exact sequence corresponding to
the short sequence \eqref{secondImsequence}, 
and consequently it collapses to the following short sequence
\begin{multline}
0 \longrightarrow S^+_{-n,n} \longrightarrow H^1(V_{-j_0-1+nk}) \longrightarrow S^-_{-n,n} 
 \longrightarrow S^-_{-n,n-1} \longrightarrow \cr
\longrightarrow H^2(V_{-j_0-1+nk}) \longrightarrow S^+_{-n,n}\longrightarrow 0
\end{multline}
Thus, we finally get the non-zero cohomologies
\begin{align}
H^1(W_{-j_0-1+nk}) &= S^+_{-n,n} \cup \{-j_0-1+nk\} \cr
H^2(W_{-j_0-1+nk}) &= S^+_{-n,n}
\end{align}
which agrees with the ansatz in \eqref{ansatz} for $m=n$. Thus, we have proved
the induction hypothesis for one case $j =-j_0-1+nk$, and all other cases can
be computed similarly. Note that we have also proven that the cohomologies
localize on singular vectors (and precisely how) in the case of integer level
and for any spin. 
We now turn to the most intricate case of fractional level $k$. 

\subsection{The cohomology for positive rational level} \label{positiverational}
We consider the fractional level $k=\frac{p}{q}$, where $p,q$ are mutually
prime strictly positive integers. We already argued that this is going to be the most
non-trivial case, since two or more singular vectors in a given Verma module imply that
$k$ is fractional. Moreover, we note that for fractional $k$, solutions
to the Kac-Kazhdan equations will only occur for
the set of spins that are multiples of $\frac{1}{2q}$. We therefore from now
on restrict to this non-trivial subset (since all other cases can be treated
more
straightforwardly).
\subsubsection*{Embedding diagrams}
For this case as well, we can algorithmically construct the embedding diagrams
for all Verma modules, using the results of Malikov \cite{Malikov} (see also
\cite{Feigin:1990ut}). The diagrams have the same ``double-string" shape as
before, but are now generated by $\rho$-centered affine Weyl reflections that depend
on the particular spin under study. We discuss in detail in the appendix
\ref{Mal} how these embedding diagrams are constructed. Although this is 
the most systematic and all-at-once procedure, below, we will follow a 
more pedestrian route. We will reconstruct the embedding diagrams
inductively simply by locating the singular vectors in a given
module. Although this approach may be more accessible and constructive, its
a posteriori justification lies in its agreement with 
the proof for the embedding diagram given in \cite{Malikov}.

\subsubsection*{The cohomology for a simple case}
To compute the cohomology, we follow an inductive procedure. First of all,
when $2j+1$ is not of the form 
\be
2j+1=r+ks \quad\hbox{such that}\quad rs > 0 \quad\hbox{or}\quad r >0,\ \, s=0 \,,
\ee 
then the module has no singular vectors and is irreducible. We have then that the Verma
module is equivalent to its dual, and is equivalent to the Wakimoto module and its
dual. 
 This is because for irreducible Verma modules, there
exists a non-degenerate Shapolov form which can be used to construct a
canonical isomorphism between the Verma module and its dual. This also implies
that the cohomology of an irreducible module 
is the same as that of its dual. 
Thus all these modules have cohomology $H^n(V_j=W_j=W^{\ast}_j)=\delta_{1,n}
\{ j \}$. This completes the calculation of the cohomology in the case of
irreducible modules.

\subsubsection*{The cohomology in the case of one singular vector}
Next, we move on to the case where the Verma module $V_{j_1}$ 
under consideration has precisely one singular vector. The values of $j_1$ for
which this is the case can be parameterized by the corresponding values of
$(r_1,s_1)$, which are related to $j_1$ by
\be
2j_1+ 1 = r_1+ks_1 \,.
\ee
When $2j_1+1$ is positive, the couple $(r_1,s_1)$ satisfies $ r_1 \in \{
1,2,\dots,p \}$ and {$s_1 \in \{0,1,\dots,q-1 \}$}. When $2j_1+1$ is negative,
we have that $r_1 \in \{ -1,-2, \dots, -p \}$ and  $s_1 \in \{-1,-2,\dots,-q
\}$. 
We have thus $2.p.q$ values of $j_1$ with precisely one singular vector.

The singular vectors are at values of the spin $j_1^{(1)}$ which are given by
 $2 j_1^{(1)} + 1 = 2j_1+1 -2r_1 = -r_1 + ks_1$. (Note that this expression can be
 of either sign.) It is now not too difficult to prove that none of the Verma
 modules associated to the singular vectors can have any singular vectors
 themselves. Thus, they are irreducible, and we have already computed their
 cohomology. We can then proceed to compute the cohomology for the case of one
 singular vector. We have the
 exact sequences:
\begin{align}
0 & \longrightarrow V_{j_1^{(1)}} \longrightarrow V_{j_1} \longrightarrow I_{j_1}
\longrightarrow 0 
\nonumber \cr
0 & \longrightarrow  I_{j_1}  \longrightarrow V^\ast_{j_1} \longrightarrow V_{j_1^{(1)}}
\longrightarrow 0 \, .
\end{align}
The first sequence indicates that modding out by the single singular vector
gives an irreducible module, while the second follows by dualizing and from
the
fact that the modules $I_{j_1}$ and $ V_{j_1^{(1)}} $ are irreducible.
Suppose that $j_1$ is positive; then $V_{j_1}=W_{j_1}$. Then we first use the second exact sequence to
compute the cohomology of $I_{j_1}$:
\begin{align}
0 & \longrightarrow H^1 (I_{j_1})  \longrightarrow \{ j_1 \} \longrightarrow \{ j_1^{(1)} \} \longrightarrow H^2 (I_{j_1})  \longrightarrow 0 \,.
\end{align}
We thus have that 
\begin{align}
 H^1 (I_{j_1}) =  \{ j_1 \} \qquad H^2 (I_{j_1}) = \{ j_1^{(1)} \}.
\end{align}
We can feed this information into the first exact sequence and we obtain:
\begin{align}
0 &  \longrightarrow  \{ j_1^{(1)} \} \longrightarrow   H^1(W_{j_1})
\longrightarrow \{j_1 \} \longrightarrow 0 \longrightarrow H^2 (W_{j_1})
\longrightarrow \{ j_1^{(1)} \} \longrightarrow 0
\end{align}
which gives us the result:
\begin{align}
H^1(W_{j_1}) =  \{j_1 \} \cup  \{ j_1^{(1)} \} \qquad H^2 (W_{j_1}) =  \{ j_1^{(1)} \} \,.
\end{align}
These are the set of the parent spin and the singular vector, and the singular vector respectively. 

For $j_1$ negative, we have that the Verma module is equivalent to the dual Wakimoto module, and repeating the analysis above, we find that 
\begin{align}
H^1(W_{j_1})= \{j_1 \} \cup  \{ j_1^{(1)} \} \qquad H^0 (W_{j_1}) =  \{ j_1^{(1)} \} \,.
\end{align}
Thus we have computed the cohomologies for all modules with a single singular
 vector.

\subsubsection*{The location of singular vectors in Verma modules}

We now turn to a classification of the cases where the parent module has
 precisely $n$ singular vectors. It is not too difficult to see that these
 cases are described as follows. Define the spin $j_n$ to be given by:
\be
\label{nsingularvectors}
2j_n+1 = r_n + ks_n \,.
\ee 
 For positive spin, we have that $j_n$ lies in the set corresponding to $r_n
 \in \{ (n-1)p+1,(n-1)p+2,\dots, n p \}$ and $s_n \in \{0,1,\dots,q-1
 \}$. When $2j_n+1$ is negative, we have that $r_n \in \{ -(n-1)p-1,-(n-1)p-2,
 \dots, - n p \}$ and $s_n \in \{-1,-2,\dots,-q \}$. We thus always have
 $2.p.q$ values of $j_n$ with precisely $n$ singular vectors. In the case of
 positive spin, which will take to be the case for our calculations, we have
 that the singular vectors
 correspond to the pairs $(r,s) \in \{ (r_n,s_n), (r_n-p,s_n+q), \dots , (r_n-(n-1)p,s_n+(n-1)q) \}$.

We also have a claim that we will treat in more detail. Suppose we
 consider the module $V_{j_n}$, and label the leading singular vectors in them as $V_{j_{n-1}^{(1)}}$ and $V_{j_{n-1}^{(n)}}$
\footnote{The notation is such that the subscript tells us the number of
  singular vectors in a given module, while the superscripts keeps track of
  its line of descent from $V_{j_n}$.}. Now list the singular vectors in these
modules. We will find below that these are in fact, the same list of
vectors. Once again, consider the leading singular vectors in this list of
$n-1$ singular vectors, and label them $V_{j_{n-2}}^{(1,1)}$ and
$V_{j_{n-2}}^{(1,n-1)}$. Consider the singular vectors in these modules; we
claim (and prove below) that the singular vectors in either of these modules
make up the remaining $n-2$ singular vectors in the original module
$V_{j_n}$. This is key to writing down short exact
sequences of Verma 
modules that will lead to computing cohomology. 

Let's prove these claims, and start with a spin $j_n$ module, associated to a module with $n$
singular vectors. We have that $2j_n+1=r_n+k s_n$ where $(r_n,s_n)$ are taken
in the set described above (and we take them to be positive for simplicity). 
The resulting spins of the modules of the singular vectors are ($l_n=1,2,\dots, n$):
\begin{align}\label{singularvectors}
2j_{n-1}^{(l_n)} + 1 &= -r_n+k s_n + 2(l_n-1) p.
\end{align}
In particular, we have the spins which we believe to be the nearest
to $j_n$ in the embedding diagram of $V_{j_n}$, namely those associated to $l_n=1,n$:
\begin{align}
2 j_{n-1}^{(1)} + 1 &= -r_n + k s_n = r_{n-1}^{(1)} + k s_{n-1}^{(1)} \cr
2 j_{n-1}^{(n)} + 1 &= -r_n + k s_n + 2(n-1)p = r_{n-1}^{(n)} + k s_{n-1}^{(n)}.
\end{align}
In the case of integer level, these appeared at an equal distance from $j_n$
 but in the rational level, this need not be the case. Modules associated to
 each of these $j$ values have $n-1$ singular vectors.
 Indeed, the associated pairs $(r_{n-1}^{(1)},s_{n-1}^{(1)})$ and $(r_{n-1}^{(n)},s_{n-1}^{(n)})$ are given by the formulae
\begin{align}
r_{n-1}^{(1)} &= p-r_n \in \{-(n-2)p, \ldots ,-(n-1)p\} \cr 
s_{n-1}^{(1)} &= s_n-q \in \{-q, \ldots , -1\} \cr
r_{n-1}^{(n)} &=  2(n-1) p - r_n \in \{(n-2)p, \ldots ,(n-1)p\} \cr 
s_{n-1}^{(n)} &=s_n \in \{1, \ldots , q \} \,.
\end{align}
Their ranges have been explicitly shown to make it clear that they indeed have
 $n-1$ singular vectors each. Let us now list the set of singular vectors in each of the modules $V_{j_{n-1}^{(1)}}$ and $V_{j_{n-1}^{(n)}}$. We will denote these respectively as $j_{n-2}^{(1,l_{n-1}^{(1)})}$ and $j_{n-2}^{(n,l_{n-1}^{(n)})}$. These satisfy 
\begin{align}
2 j_{n-2}^{(1,l_{n-1}^{(1)})} + 1 &= -r_{n}^{(1)} + k s_{n}^{(1)} - 2 (l_{n-1}^{(1)}-1) p \quad \hbox{with} \quad l_{n-1}^{(1)} \in \{1\ldots n-1 \} \cr
&= r_n-p + k (s_n-q) - 2 (  l_{n-1}^{(1)} -1) p \cr
&= r_n + k s_n - 2 l_{n-1}^{(1)} p  
\end{align}
and
\begin{align}\label{secondsingular}
2 j_{n-2}^{(n,l_{(n-1)}^{(n)})} +1 &= -r_{n-1}^{(n)} + k s_{n-1}^{(n)} + 2 (l^{(n)}_{n-1}-1) p \quad \hbox{with} \quad l_{n-1}^{(n)} \in \{1\ldots n-1 \}\cr
&= r_n-2(n-1)p+ k s_n + 2  (l^{(n)}_{n-1}-1) p\cr
&= r_n+k s_n - 2 (n - l^{(n)}_{n-1}) p
\end{align}
It is clear from \eqref{secondsingular} that the singular vectors in each
 of the two lists is identical as each of the $l_{n-1}$'s go from $1$ to $n-1$. In particular, we see that 
\be
j_{n-2}^{(1,l_{n-1}^{(1)})} = j_{n-2}^{(n,n-l_{(n-1)}^{(n)})} \,.
\ee
Without loss of generality, we may therefore safely omit the first superscript and denote them as $j_{n-2}^{(1)}\ldots j_{n-2}^{(n-1)}$. Let us focus on either of the leading singular vectors, which have $l_{n-1} = 1, n-1$. These have the $r_{n-2}$ and $s_{n-2}$ values given by 
\begin{align}
r_{n-2}^{(1)} &= r_n - 2p \in \{(n-3)p, \ldots ,(n-2)p\} \cr
s_{n-2}^{(1)} &= s_n \in \{1, \ldots, q\} \cr
r_{n-2}^{(n-1)} &= r_n-(2n-3)p \in \{-(n-2)p, \ldots, -(n-3)p \} \cr
s_{n-2}^{(n-1)} &= s_n -q \in \{-q, \ldots, -1\} \,.
\end{align}
It is clear that their $(r_{n-2},s_{n-2})$ values are such that the modules
  $V_{j_{n-2}^{(1)}}$ and $V_{j_{n-2}^{(n-1)}}$ have $n-2$ singular
  vectors each, which we label $j_{n-3}^{(1, m^{(1)}_{n-2} )}$ and
  $j_{n-3}^{(n-1, m^{(n-1)}_{n-2} )}$ respectively. These satisfy the equations 
\begin{align}
2 j_{n-3}^{(1, m^{(1)}_{n-2} )} + 1 &= -r^{(1)}_{n-2} + k s^{(1)}_{n-2} + 2(m^{(1)}_{n-2}-1)p \cr
&= - (r_n- 2 p) + k s_n +2(m^{(1)}_{n-2}-1)p\cr 
&= -r_n+k s_n + 2 m^{(1)} p 
\end{align}
and
\begin{align}
2 j_{n-3}^{(n-1, m^{(n-1)}_{n-2})} +1 &= -r^{(n-1)}_{n-2} + k s^{(n-1)}_{n-2} - 2(m^{(n-1)}_{n-2}-1)p \cr
&= (2n-3)p - r_n + ks_n - p - 2(m^{(n-1)}_{n-2}-1)p \cr
&= -r_n + k s_n +2 ((n-1)-m_{n-2}^{(n-1)}) \,. 
\end{align}
where $m^{(1)}_{n-2}$ and $m^{(n-1)}_{n-2}$ take values in $\{1,\ldots ,
 n-2\}$.
 Once again, we see that both lists of singular vectors coincide if we identify 
\be
 j_{n-3}^{(1,m^{(1)}_{n-2} )} =  j_{n-3}^{(n-1,(n-1)-m^{(n-1)}_{n-2})} \,.
\ee
Once again, the first superscript index is redundant and we may denote them as $j_{n-3}^{(1)}\ldots j_{n-3}^{(n-2)}$. Moreover, by comparing with \eqref{singularvectors}, we observe that the singular vectors in both $V_{j_{n-2}^{(1)}}$ and $V_{j_{n-2}^{(n-1)}}$ (the modules corresponding to the leading second generation singular vectors) precisely account for all but two of the singular vectors in the original module $V_{j_n}$. We have thus proved the assertion made at the beginning of this section.

\subsubsection*{Proof by induction}
We now make an educated guess for the embedding diagram of these
modules. As already remarked, the guess is exact, as proven in \cite{Malikov}
(see also our appendix \ref{Mal})\footnote{Again, in exceptional cases,
linear embedding diagrams arise. The calculation of the cohomology is then
easily adapted from the appendix to \cite{Mukhi} and
our subsection on the linear case for integer $k$.}:
\begin{equation}\label{embedexfrac}
\xymatrix{
& V_{j_{n-1}^{(1)} }\ar[ddr]\ar[r]& V_{ j_{n-2}^{(n-1)} }& V_{j_2^{(1)} }
\ar[ddr] \ar[r]  & V_{j_1^{(2) }}  \ar[dr] & \\
V_{j_n} \ar[ur]\ar[dr]&& \ldots&   \ldots& & V_{j_0} \\
& V_{j_{n-1}^{(n)} }\ar[uur]\ar[r]& V_{j_{n-2}^{(1)}    }& V_{j_2^{(3)  } } 
\ar[uur]\ar[r]   & V_{j_1^{(1)}} \ar[ur] & 
\\
}
\end{equation}
 The knowledge of the embedding diagram allows us to write down the exact sequences:
\begin{multline}\label{firstImageseq}
0\longrightarrow \hbox{Im}(V_{j_{n-2}^{(1)}}\oplus V_{j_{n-2}^{(n-1)}}) \longrightarrow V_{j_{n-1}^{(1)}}\oplus V_{j_{n-1}^{(n)} }\longrightarrow 
 \hbox{Im}(V_{j_{n-1}^{(1)}}\oplus V_{j_{n-1}^{(n)} })\longrightarrow 0
\end{multline}
\begin{equation}\label{secondImageseq}
0\longrightarrow \hbox{Im}(V_{j_{n-1}^{(1)}}\oplus V_{j_{n-1}^{(n)} })\longrightarrow V_{j_n}\longrightarrow I_{j_n}\longrightarrow 0
\end{equation}  
The second sequence codes the irreducibility of the module obtained by modding
out by the leading singular vectors, while the first one encodes the double
counting
in this modding out.
{From} here on, the proof for the cohomology of the BRST operator $Q$ is identical to the
integer level case and we once again find that cohomology elements are given
by the singular vectors. As for the integer level case, we need to postulate
not only the cohomology of the Wakimoto modules but also those of the image
modules appearing in \eqref{firstImageseq}. We make the following induction hypothesis for all $m$ values in the range $0 \le m \le n-1$ : 
\begin{align}\label{pbyqansatz}
&\hbox{For $j_m > -\half$ }&H^1(W_{j_m} = V_{j_m}) &= \{j_{m-1}^{(1)} \ldots j_{m-1}^{(m)} \} \cup \{j_m\} \cr
& & H^2(W_{j_m} = V_{j_m}) &= \{j_{m-1}^{(1)} \ldots j_{m-1}^{(m)} \}\cr
&\hbox{For $j_m < -\half$} &H^1(W_{j_m} = V^{*}_{j_m}) &= \{j_{m-1}^{(1)} \ldots j_{m-1}^{(m)} \} \cup \{j_m\}  \cr 
& &H^0(W_{j_m} = V^{*}_{j_m}) &= \{j_{m-1}^{(1)} \ldots j_{m-1}^{(m)} \} \cr
& &H^q(W^*_{j_m}) &= \delta_{q,1}\ \{j_m\}\quad  \forall j_m
\end{align}
while for the image modules appearing in \eqref{firstImageseq} and \eqref{secondImageseq}, we postulate
\begin{align}
H^1(\hbox{Im}(V_{j_{n-2}^{(1)}} \oplus V_{j_{n-2}^{(n-1)} } )) & = \{j_{n-2}^{(1)} \ldots j_{n-2}^{(n-1)} \} \cr
H^2(\hbox{Im}(V_{j_{n-2}^{(1)}} \oplus V_{j_{n-2}^{(n-1)} } )) & = \{j_{n-3}^{(1)} \ldots j_{n-3}^{(n-2)} \} \cr
& = \{j_{n-1}^{(2)} \ldots j_{n-1}^{(n-1)} \}
\end{align}
where we have generalized \eqref{imageone} and \eqref{imagetwo}. We will try to prove \eqref{pbyqansatz} for $m=n$. We begin by using our induction hypothesis in the long exact sequence corresponding to the short  sequence in equation \eqref{firstImageseq}. We obtain the sequence
\begin{multline}
0 \longrightarrow H^0(V_{j_{n-1}^{(1)}}\oplus V_{j_{n-1}^{(n)} }) \longrightarrow 
\{j_{n-2}^{(1)} \ldots j_{n-2}^{(n-1)} \} \longrightarrow 
\{j_{n-1}^{(1)}, j_{n-2}^{(1)} \ldots j_{n-2}^{(n-1)} \} \longrightarrow \cr
\longrightarrow
H^1(\hbox{Im}(V_{j_{n-1}^{(1)}}\oplus V_{j_{n-1}^{(n)} })
\longrightarrow  \{j_{n-1}^{(2)} \ldots j_{n-1}^{(n-1)} \} 
\longrightarrow H^2(V_{j_{n-1}^{(1)}}\oplus V_{j_{n-1}^{(n)} }) \longrightarrow 0
\end{multline}
from which we solve for the cohomology of the image modules : 
\begin{align}
H^1(\hbox{Im}(V_{j_{n-1}^{(1)}}\oplus V_{j_{n-1}^{(n)} }) &= \{j_{n-1}^{(1)} \ldots j_{n-1}^{(n)} \}  \cr
H^2(\hbox{Im}(V_{j_{n-1}^{(1)}}\oplus V_{j_{n-1}^{(n)} }) &= \{j_{n-2}^{(1)}
\ldots j_{n-2}^{(n-1)} \} .
\end{align}
Thus, we have inductively verified the ansatz for the cohomology of the image
modules. The cohomology for the dual of the image is computed similarly, and
one obtains the same cohomology with the ghost number zero $H^0$ and
ghost number two cohomology $H^2$ interchanged. 
We now plug this result into the long exact sequence corresponding to the dual
of the short exact
 sequence in \eqref{secondImageseq} to get
\begin{multline}
0 \longrightarrow H^0(I_{j_n}) \longrightarrow  0 \longrightarrow  \{j_{n-2}^{(1)} \ldots j_{n-2}^{(n-1)} \} \longrightarrow H^1(I_{j_n}) \longrightarrow \cr
\longrightarrow \{j_n\} \longrightarrow \{j_{n-1}^{(1)} \ldots j_{n-1}^{(n)} \}  \longrightarrow H^2(I_{j_n}) \longrightarrow 0
\end{multline}
leading to
\begin{align}
H^1(I_{j_n}) &= \{j_n \} \cup \{ j_{n-2}^{(1)} \ldots j_{n-2}^{(n-1)} \} \cr
H^2(I_{j_n}) &= \{j_{n-1}^{(1)} \ldots j_{n-1}^{(n)} \} \,.
\end{align}
Substituting all these results into the long exact sequence corresponding to \eqref{secondImageseq}, we find
\begin{multline}
0 \longrightarrow H^0(W_{j_n}) \longrightarrow 0 \longrightarrow \{j_{n-1}^{(1)} \ldots j_{n-1}^{(n)} \} \longrightarrow H^1(W_{j_n})
\longrightarrow \{j_n, j_{n-2}^{(1)} \ldots j_{n-2}^{(n-1)} \}  \longrightarrow \cr
\longrightarrow \{j_{n-2}^{(1)} \ldots j_{n-2}^{(n-1)} \} 
\longrightarrow H^2(W_{j_n}) \longrightarrow \{j_{n-1}^{(1)} \ldots j_{n-1}^{(n)} \} \longrightarrow 0
\end{multline} 
from which we conclude
\begin{align}
H^1(W_{j_n}) & = \{j_n \}  \cup \{j_{n-1}^{(1)} \ldots j_{n-1}^{(n)} \} \cr
H^2(W_{j_n}) & = \{j_{n-1}^{(1)} \ldots j_{n-1}^{(n)} \} 
\end{align}
By comparing with equation \eqref{pbyqansatz}, we observe that we have derived the
result for $W_{j_n}$ by assuming the result for all $m <n$. This is our answer
for the cohomology of Wakimoto modules and thus the complete set of
topological observables for the level $k=\frac{p}{q}$ twisted $SL(2,\IR)/U(1)$ 
model.  

\subsection{Alternative proofs}
An alternative inductive proof for the cohomology of the Wakimoto modules is
 possible, based on another pair of short exact sequences\footnote{We would
 like to thank Edward Frenkel for indicating to us the idea behind this
 alternative proof.}. Again, for simplicity only, we restrict to the case of
 positive $2j_n+1=r_n + k s_n$. The induction hypothesis consists of the
 validity of the cohomology that we found above for modules with less than $n$
 singular vectors in \eqref{pbyqansatz}, and, of the cohomology of the
 corresponding irreducible modules. (The latter cohomology depends on the sign
 of the spin.) Now, we can construct a first short exact sequence by modding
 out the Verma module of spin $j_n$ by one of its two leading singular
 vectors, say $j_{n-1}^{(1)}$, to
 produce a new module $X_n$:
\begin{align}
0 & \rightarrow V_{j_{n-1}^{(1)}} \rightarrow V_{j_n} \rightarrow X_n
\rightarrow 0
\nonumber \\
0 & \rightarrow I_{j_{n-1}^{(n)}} \rightarrow X_n \rightarrow I_{j_n}
\rightarrow 0.
\end{align}
We then used the module $X_n$ in a second sequence which says that after
eliminating
the further singular vector $j_{n-1}^{(n)}$ (and none of its descendants since
they have already been modded out), we obtain an irreducible module. Using
these two short exact sequences and their duals it is not hard to show that
the induction step can be proven (both on the cohomology of the irreducible
module and on the cohomology of the Wakimoto modules). For the reader's
convenience we record an intermediate result, namely the cohomology of the
modules $X_n$ (for $j_n$ positive):
\begin{align}
H^1 (X_n) &=  \{ j_n \} \cup \{ j_{n-2}^{(1)}, \dots, j_{n-2}^{(n-1)} \} \cr
H^2 (X_n) &=  \{ j_{n-1}^{(1)}\} \cup \{ j_{n-2}^{(1)}, \dots, j_{n-2}^{(n-1)} \} \,.
\end{align}
Furthermore, from the first alternative proof follows a second. We could
choose to mod out first by the other leading singular vector. We obtain the
short exact sequences:
\begin{align}
0 & \rightarrow V_{j_{n-1}^{(n)}} \rightarrow V_{j_n} \rightarrow Y_n
\rightarrow 0
\nonumber \\
0 & \rightarrow I_{j_{n-1}^{(1)}} \rightarrow X_n \rightarrow I_{j_n}
\rightarrow 0.
\end{align}
The proof follows familiar lines. We quote the intermediate result for the
cohomology
of the new modules $Y_n$:
\begin{align}
H^0 (Y_n) &= \{ j_{n-2}^{(1)}, \dots, j_{n-2}^{(n-1)} \}
\cr
H^1 (Y_n) &=\{ j_n\} \cup\{ j_{n-1}^{(1)},\ldots ,  j_{n-1}^{(n-1)}\} \cup \{j_{n-2}^{(1)}, \dots, j_{n-2}^{(n-1)}\}
\cr
H^2 (Y_n) &= \{ j_{n-1}^{(1)}, \dots, j_{n-1}^{(n)}\}.
\end{align}
Note that these modules have cohomology at three different ghost
 numbers.\footnote{Algebraically, it is intuitive that the two leading singular
 vectors give rise to proofs with slightly different structure, since only one
 of the two leading singular spins is related to the parent spin by an
 elementary Weyl reflection, as defined and discussed in
 the appendix \ref{Mal}.}

\section{Concluding remarks}
Our main result is that the cohomology of the topologically twisted coset theory is given in terms of the singular vectors of the Wakimoto modules constructed over the $SL(2,\IR)$ current algebra. The precise form in which this is realized can be seen in equation \eqref{pbyqansatz}. We conclude by making a few remarks on further avenues to explore, and on the uses of the rather technical results we have obtained regarding the topological observables of the cigar.
\begin{itemize}
\item It is possible to give explicit representatives of the cohomology, using the formulae obtained for singular vectors in $SL(2,\IR)$ modules \cite{MFF}\cite{Bauer:1993jj}. Following the analysis in \cite{Mukhi}, rewriting these cohomology elements in Wakimoto variables could be very useful in establishing the relation between twisted cosets and bosonic string theories.
\item It would be instructive to further analyze, as was done in  \cite{Ashok:2005xc}
  for the particular case of the topological cigar at level $k=1$ (i.e. the
  conifold), how the operation of spectral flow acts on the observables of the
  topological cigar. 
\item The relationship between the statement that the cohomology localizes on
  singular vectors and the statement \cite{Feigin:1997ha} that $N=2$ topological singular vectors
always arise (through the Kazama-Suzuki map) from $SL(2,\IR)$ singular vectors are
  tantalizingly close. It is fairly clear that 
clarifying the details of the relation will also
  illuminate the role of spectral flow. 
\item An important application we have in mind for these results is in the
  context
of the $N=2$ topological string (see e.g. \cite{Bershadsky:1993cx} for a
  review). Now that we obtained the observables on the topological cigar,
the algorithm to obtain the closed string spectrum of topological strings on
  non-compact Gepner models consisting of (twisted) $N=2$ minimal models and topological
  cigars is clear. Indeed, the chiral cohomology on the $N=2$ minimal models is
  straightforward to compute, and is well-known. (When working with a unitary spectrum, the
  powerful results of \cite{Lerche:1989uy} make the calculation much easier.)
 Furthermore, one could then combine left- and right-movers to form a consistent
  spectrum, following Gepner's technique of GSO
  projecting onto integer $U(1)_R$ charges. In the present context, the GSO
  projection will boil down to a simple orbifold by an abelian group that will
  depend only on the levels involved in the non-compact Gepner model. (This is
  clear from the form of the $N=2$ $U(1)_R$ current in these models and is for
  instance discussed in  
\cite{Eguchi:2000tc}\cite{Murthy:2003es}\cite{Eguchi:2004yi}\cite{Israel:2004ir}\cite{Eguchi:2004ik}, 
following Gepner.) It will then be very interesting to combine these
  observations together with the fact that the $N=2$ topological string basically
  corresponds to a bosonic string theory (see
  e.g. \cite{Bershadsky:1993cx}). This would be 
a worldsheet approach to isolate integrable subsectors of string theories
  on
 (non-compact) Calabi-Yau manifolds \cite{Aganagic}.
\end{itemize}

\section*{Acknowledgements}

We are grateful to Wendy Lowen for helpful correspondence and to Edward Frenkel for insightful comments on these problems and on the appendix to \cite{Mukhi}. 
We are also thankful to Eleonora Dell'Aquila, Jaume Gomis and Sameer Murthy for discussions and comments on the draft. 
Research of S.A at the Perimeter Institute is supported in part by funds from NSERC of Canada and by  MEDT of Ontario, and research of J.T. is partly supported by  EU-contract MRTN-CT-2004-05104.
\begin{appendix}
\section{A few notes on affine algebra}
\label{Mal}
The calculation of the cohomology performed in the bulk of the paper heavily
leans on an understanding of the embedding structure of Verma modules for the
rank two Kac-Moody algebra $sl_2$. In this appendix, we briefly review some of
the mathematics involved in understanding the embedding diagrams. We hope this
will make the relevant mathematics literature more readable to the interested
physicist. 

We introduce some notation and refer to standard text books on affine algebras
for further information on the following standard concepts. We introduce the
algebra $sl_2$, where the notation is meant to indicate that we consider the
algebra over the complex numbers. We follow (in this appendix only)
$su(2)$ conventions for the
structure constants and metric of the corresponding affine algebra (see e.g.
\cite{DiFrancesco:1997nk} for standard conventions). The affine
algebra has simple positive roots $\alpha_1$ (the simple positive root of the
ordinary, horizontal $su(2)$ subalgebra) and $\alpha_0= \delta-\alpha_1$, where $\delta$ is the
imaginary root. We take the algebra to be at (bosonic) level $k_{su(2)}$.

We can define $\rho$-centered affine Weyl reflections (where $\rho$ is defined
in the usual way, as having inner product equal to one with all positive
simple roots). These $\rho$-centered affine Weyl reflections $s_\alpha^\rho$ 
with respect to
the root $\alpha$ are given in terms of the affine Weyl reflections 
$s_\alpha$ by the formula:
\begin{align}
s_{\alpha} \hat{\lambda} &= \hat{\lambda} -
\frac{2(\alpha,\hat{\lambda})}{(\alpha,\alpha)}
\nonumber \\
s_{ \alpha }^{\rho}  ( \lambda ) &= s_\alpha (\lambda+ \rho)-\rho.
\end{align}
We will mostly refer only to the action of the $\rho$-centered affine Weyl
reflections on the part of the weight that lies in the direction of the
horizontal $su(2)$ subalgebra of the full Kac-Moody algebra (i.e. the standard
spin of the base representation). On the spin $j$ of a representation
(corresponding to a horizontal weight $\lambda_1=j \alpha_1$, these
Weyl reflections act as follows:
\begin{align}
s^{\rho}_{\alpha_1+l \delta} (j) &= -l(k_{su(2)}+2)-j-1
\nonumber \\
s^\rho_{ -\alpha_1+l \delta  } (j) &= l(k_{su(2)}+2)-j-1.
\end{align}
Following \cite{Malikov}
we also define an operation $S$ on the spin and the level of the affine
algebra
which acts as $(j,k_{su(2)}) \mapsto (-j-1,-k_{su(2)}-4)$.

\subsection{Flipping}
The statements on the structure of the embedding of Verma modules derived in
\cite{Malikov}, and restricted to the case of relevance here, can now be made
more transparent, using the above notations. 

First of all, the main theorem on the embedding structure in \cite{Malikov}, 
in paragraph 2, case A, pertains to the case where the level of the
$sl_2$ algebra, which we have named $k_{su(2)}$ is larger than $-2$. However,
the case of interest to us in the bulk can be reduced to this case via the
operation $S$. Indeed, consider a spin and bosonic $su(2)$ level $(j_n,-k-2)$ as in the bulk of
our paper. (Note that $k_{su(2)}= -k-2$, where $k$ is the supersymmetric
$SL(2,R)$ level used in the bulk of the paper.) By acting with the operation $S$, we find the spin and level
$(-j_n-1,k-2)$. Since our supersymmetric level $k$ is larger than zero, we do
find that the new bosonic level $k-2$ is larger than $-2$, and that we now
therefore fall into the framework of the embedding diagrams discussed in
paragraph 2, case A of \cite{Malikov}.

\subsection{Constructing the embedding diagram}
The results of Malikov \cite{Malikov} then permit us to construct the
embedding diagrams for all cases. In particular, we concentrate on the most
difficult case of $k=p/q$ fractional (with $p,q$ both strictly positive
mutually
prime integers). First note that the Kaz-Kazhdan equation can
be written more invariantly as being satisfied for an affine weight $\lambda$
when:
\begin{align}
2(\lambda+ \rho, \alpha) &= m (\alpha, \alpha)
\end{align}
for some positive root $\alpha$ and $m \in N_0$ (i.e. $m$ a strictly positive
integer). This translates directly into the equation used in the bulk of the
paper. However, following \cite{Malikov}, we will now consider solutions to
this equation for $m$ any integer. In other words, we will study solutions to
the equation:
\begin{align}
-2j_n-1 = -m - k n
\end{align}
which is the Kaz-Kazhdan equation for the flipped spin defined above, and with
respect to the flipped supersymmetric level $k$, and this for any integers $m$
and $n$. Pick now the smallest positive integer $n$ such that the equation is
satisfied, and the smallest strictly negative integer $n$ such that the
equation holds. For the values $j_n$ defined in the bulk of the paper,
in equation (\ref{nsingularvectors}), it is not hard to see that these values correspond to 
$s_n, s_n-q$ and $s_n+q,s_n$ for $j_n$ positive and negative respectively.
These, via the Kac-Kazhdan equations above correspond to positive affine roots
which are, for $2j_n+1$ positive:
\begin{align}
\alpha_1 + s_n \delta
\nonumber \\
-\alpha_1 + (q-s_n) \delta
\end{align}
while for $2j_n+1$ negative, we consider the positive roots:
\begin{align}
\alpha_1 + (q+s_n) \delta
\nonumber \\
-\alpha_1 -s_n \delta.
\end{align}
The results of \cite{Malikov} show that the embedding diagrams of the Verma
module with spin $j$ can be constructed from an initial Verma module $j_0$ by
$\rho$-centered Weyl-reflecting from this initial point, and adding a node to the
embedding diagram for each step on the way to the destination $j$. The node
$j_0$ corresponds to the unique spin satisfying the equations:
\begin{align}
0 \le -2 j_0-1 + s_n k \le p
\end{align}
respectively
\begin{align}
0 \le 2j_0+1 - s_n k \le p,
\end{align}
and lying in the orbit of $j$ under the above $\rho$-centered Weyl reflections.

Since this prescription is rather abstract (although already more concrete
than the original one \cite{Malikov}), let us first illustrate this in the
case  where the level $k$ is integer, and see how we recuperate the diagrams
used in the bulk of the paper.

\subsubsection*{The case of $k$ integer}
For $k$ integer, we have that $q=1$, and that for $2j_n+1$ positive, say,
we have that $s_n=0$. Consequently, the allowed $\rho$-centered Weyl
reflections are those with respect to $\alpha_1$ and $\alpha_1+\delta$. These
are the standard $\rho$-centered Weyl reflections, and those are precisely the
ones we used to construct the diagrams in the case of integer level. Moreover,
the condition on the final spin becomes $0 \le -2j_0-1 \le k=p$ which is
precisely what we found in the bulk of the paper.

\subsubsection*{The general case}
In the general case too, one can show with a bit more tedious but
straightforward calculations that the above systematic construction
of the embedding diagrams agrees on the nose with the analysis in terms of
singular vectors only performed in the bulk of the paper. In fact, this
justifies the approach in the bulk of the paper, via the detailed proof of
Malikov of the embedding structure of all Verma modules for the affine $sl_2$ algebra.

\section{A case study}

As we have seen, for $j$-values that are not of the form $2j+1 \in k\IZ$, the
sub-module structure is not linear. The simplest example of a non-linear
embedding of sub-modules occurs when the embedding diagram has only four
entries (the ``rhombus" diagram). 
From the general embedding diagram \eqref{embedex} for integer $k$, we see
that this happens
 for $j = j_0 \pm k$. Let us discuss each of these cases in turn and demonstrate explicitly how one would compute cohomology inductively. 

\subsection{The case $j = j_0+k$}
For $j > -\half$, the Wakimoto modules are isomorphic to the Verma modules and
 we can read
 off the submodule structure from \cite{Malikov} to be  (with $W_{j}=V_{j}$)
\begin{equation}\label{rhombus}
\xymatrix{
&V_{-j_0-1}\ar[dr]&\\
V_{j_0+k} \ar[dr]\ar[ur] & & V_{j_0} \\
& V_{-j_0-k-1}\ar[ur] 
}       
\end{equation}
In order to find the cohomology of Q in this module $W_{j}$, we have to first find the cohomology of the modules $W_{-j_0-1}$ and $W_{-j_0-k-1}$. However, these have linear sub-module structure and their cohomology can be computed using the techniques of Frenkel in \cite{Mukhi}. Let us illustrate this in the case of $W_{-j_0-1}=V_{-j_0-1}$, which has the embedding structure
\be
V_{-j_0-1}\longrightarrow V_{j_0} \,.
\ee
Now, it follows that the following sequence of modules is a short exact sequence :
\be
0\longrightarrow I_{-j_0-1}\longrightarrow W^*_{-j_0-1}\longrightarrow W_{j_0} \longrightarrow 0 \,.
\ee
So far we have only focused on the modules over the $SL(2,\IR)$ current algebra. In order to compute the cohomology of the Q operator in \eqref{Q}, we have to tensor the $SL(2,\IR)$ modules with the Fock modules over the $(b,c)$ and $X$ algebra. Although we will suppress these tensor products while writing out the sequences of modules, it is good to keep in mind that the states (observables) we will find as cohomology elements belong to the Hilbert space in \eqref{hilbert}. 

The corresponding long exact sequence of cohomology is given by 
\be
\ldots \longrightarrow H^q(I_{-j_0-1})\longrightarrow H^q(W^*_{-j_0-1})\longrightarrow H^q(W_{j_0})\longrightarrow H^{q+1}(I_{-j_0-1}) \longrightarrow \ldots
\ee
Using \eqref{dualwaki} and \eqref{theorem}, our long exact sequence collapses to the following short exact sequences
\begin{align}
0\longrightarrow H^1(I_{-j_0-1})\longrightarrow \{-j_0-1\} \longrightarrow 0 \cr
0\longrightarrow \{j_0\}\longrightarrow H^2(I_{-j_0-1})\longrightarrow 0 
\end{align}
all other cohomology groups being trivial. We thus get $H^1(I_{-j_0-1})=\{-j_0-1\}$ and $H^2(I_{-j_0-1})=\{j_0 \}$. Now consider the short exact sequence dual to the one considered above
\be
0\longrightarrow  W^*_{j_0}\longrightarrow W_{-j_0-1}\longrightarrow I_{-j_0-1} \longrightarrow 0 \,.
\ee
Once again, considering the corresponding long exact sequence and using the results derived so far, we get the collapsed short exact sequences for the cohomology
\begin{align}
0\longrightarrow \{j_0\}\longrightarrow H^1(W_{-j_0-1})\longrightarrow\{-j_0-1\}\longrightarrow 0 \cr
0\longrightarrow H^1(W_{-j_0-1})\longrightarrow\{j_0\} \longrightarrow 0 
\end{align}
Similar manipulations can be done for $j=-k-j_0-1$ and we get the following result for the cohomology :
\begin{align}
\hbox{For}\ j=-j_0-1,\ \quad &H^1(W_{-j_0-1})=\{-j_0-1,j_0\}\cr
& H^2(W_{-j_0-1})=\{j_0\} \cr
\hbox{For}\ j=-k-j_0-1,\ \quad &H^1(W_{-k-j_0-1})=\{-k-j_0-1,j_0\}\cr 
&H^0(W_{-k-j_0-1})=\{j_0\}
\end{align}
We now have the ingredients necessary to compute the cohomology of $W_{j_0+k}$ whose submodule structure is shown in \eqref{rhombus}. We claim that the following sequences are exact
\begin{align}
0\longrightarrow I_{j_0}\longrightarrow V_{-k-j_0-1}\oplus V_{-j_0-1} \longrightarrow \hbox{Im}(V_{-k-j_0-1}\oplus V_{-j_0-1})\longrightarrow 0 \label{first}\\
0\longrightarrow \hbox{Im}(V_{-k-j_0-1}\oplus V_{-j_0-1})\longrightarrow V_{j_0+k}\longrightarrow I_{j_0+k} \longrightarrow 0 \label{second}
\end{align}
In all, there are four unknown cohomologies, those of the modules $V_{j_0+k}$,
$I_{j_0+k}$, and the cohomology 
of the image module that appears in sequence \eqref{first} and its dual. 
Using these two short exact sequences and their dual sequences, it is possible to compute the cohomology of all four modules.

We first solve for the cohomology of Im$(V_{-k-j_0-1}\oplus V_{-j_0-1})$. Using the fact that 
\be
V_{-j_0-k-1} = W^*_{-j_0-k-1}\quad \hbox{and} \quad V_{-j_0-1}= W_{-j_0-1}\,, 
\ee
the long exact sequence corresponding to the short exact sequence in \eqref{first} collapses to 
\begin{multline}
0\longrightarrow H^0(\hbox{Im}(V_{-k-j_0-1}\oplus V_{-j_0-1}))\longrightarrow \{j_0\} \longrightarrow 
\{-k-j_0-1, -j_0-1, j_0\} \longrightarrow\cr 
\longrightarrow H^1(\hbox{Im}(V_{-k-j_0-1}\oplus V_{-j_0-1}))\longrightarrow 0 
\end{multline}
\begin{align}
0\longrightarrow \{j_0\} \longrightarrow H^2(\hbox{Im}(V_{-k-j_0-1}\oplus V_{-j_0-1}))\longrightarrow0
\end{align}
This leads to 
\begin{align}
H^1(\hbox{Im}(V_{-k-j_0-1}\oplus V_{-j_0-1})) &=\{-k-j_0-1,-j_0-1\} \cr
H^2(\hbox{Im}(V_{-k-j_0-1}\oplus V_{-j_0-1})) &=\{j_0\} \,. 
\end{align}
Let us now consider  the sequence dual to \eqref{first}. Using similar techniques, we find that the dual of the image module has the same cohomology, with $H^0$ replaced by $H^2$. Let us now apply these results to the  dual of the exact sequence in \eqref{second}:
\be
0\longrightarrow I_{j_0+k} \longrightarrow V^*_{j_0+k}\longrightarrow \hbox{Im}^*(V_{-k-j_0-1}\oplus V_{-j_0-1}) \longrightarrow 0 \,.
\ee
Since, $V_{j=j_0+k}=W_{j_0+k}$, the corresponding long exact sequence collapses to 
\begin{align}
0 \longrightarrow \{j_0\} \longrightarrow H^1(I_{j_0+k})\longrightarrow \{j_0+k\}\longrightarrow 0 \cr
0 \longrightarrow \{-k-j_0-1, -j_0-1\} \longrightarrow H^2(I_{j_0+k})\longrightarrow 0
\end{align}
from which we get $H^1(I_{j_0+k})=\{j_0,j_0+k\}$ and $H^2(I_{j_0+k})=\{-k-j_0-1,-j_0-1\}$. Now, writing out the long exact sequence corresponding to the short exact sequence in \eqref{second} and using all the results obtained so far, we get the following non-trivial long exact sequence 
\begin{multline}
0 \longrightarrow\{-k-j_0-1, -j_0-1\} \longrightarrow H^1(W_{j_0+k}) \longrightarrow \{j_0, j_0+k\}\longrightarrow \{j_0\}\longrightarrow \\
\longrightarrow H^2(W_{j_0+k}) \longrightarrow \{-k-j_0-1, -j_0-1\} \longrightarrow 0
\end{multline}
The long sequence breaks as shown in the equation into two subsequences, which can be solved to obtain
\begin{align}
H^1(W_{j_0+k}) &=\{-k-j_0-1,-j_0-1,j_0+k\} \cr 
H^2(W_{j_0+k}) &=\{-k-j_0-1, -j_0-1\}
\end{align}

\subsection{The case $j=j_0-k$}

As mentioned at the beginning of this section, there is another value of $j$
for which the sub-module structure is the same as in \eqref{rhombus}, $j=j_0-k
< 0$. However, since for negative $j$, Wakimoto modules are isomorphic to dual
Verma modules \cite{Mukhi}, the directions in the arrows of the embedding
diagram
for the Wakimoto module are opposite to what
was considered 
earlier  (since $W_{j_0-k} = V^*_{j_0-k}$): 
\begin{equation}\label{rhombustwo}
\xymatrix{
&V^*_{-j_0-1}\ar[dl]&\\
V^*_{j_0-k} & & V^*_{j_0}\ar[ul]\ar[dl]\\
&V^*_{-k-j_0-1}\ar[ul]
}       
\end{equation}
Since this case is different from the $j > 0$ case, let us go through the
exercise of computing cohomology explicitly. The short exact sequences in
\eqref{first} and \eqref{second} remain the same, with $j_0+k$ replaced by
$j_0-k$. However, when writing out the long exact sequence of cohomologies, it
is important to use the equality 
$V_{j_0-k} = W^*_{j_0-k}$. With this in mind, let us proceed as before. 

The computation of the cohomology of Im$(V_{-k-j_0-1}\oplus V_{-j_0-1})$ using the sequence in \eqref{second} remains unchanged. We are thus left with the sequence
\be\label{shortninebytwo}
0\longrightarrow \hbox{Im}(V_{-k-j_0-1}\oplus
 V_{-j_0-1})\longrightarrow W^*_{j_0-k}\longrightarrow I_{j_0-k} \longrightarrow 0 \,.
\ee 
The corresponding long exact sequence breaks up into 
\begin{align}
0\longrightarrow H^0(I_{j_0-k})\longrightarrow \{-k-j_0-1,-j_0-1 \}\longrightarrow 0 \cr
0\longrightarrow \{j_0-k\}\longrightarrow H^1(I_{j_0-k})\longrightarrow \{j_0,-j_0-1\}\longrightarrow 0
\end{align}
leading to $H^0(I_{j_0-k})=\{-k-j_0-1,-j_0-1\}$ and $H^1(I_{j_0-k})=\{j_0-k,j_0\}$. Consider now the sequence dual to \eqref{shortninebytwo}
\be
0\longrightarrow I_{j_0-k} \longrightarrow W_{j_0-k}\longrightarrow 
 \hbox{Im}^*(V_{-k-j_0-1}\oplus V_{-j_0-1})\longrightarrow 0 \,.
\ee 
The long exact sequence is of the form
\begin{multline}
0\longrightarrow \{-k-j_0-1,-j_0-1\}\longrightarrow H^0(W_{j_0-k})\longrightarrow 
\{j_0\} \longrightarrow \cr
\longrightarrow \{j_0,j_0-k\}
\longrightarrow  H^1(W_{j_0-k})\longrightarrow \{-k-j_0-1,-j_0-1\}\longrightarrow 0
\end{multline}
which breaks into two shorter sequences (as indicated in the diagram). Solving for the cohomology groups, we get
\begin{align}
H^1(W_{j_0-k}) &=\{j_0-k,-k-j_0-1,-j_0-1\} \cr 
H^0(W_{j_0-k}) &= \{-k-j_0-1,-j_0-1\}
\end{align}
The notable feature of this analysis is the following : 
\begin{itemize}
\item The spin $j$-values of the cohomology elements correspond to the
  singular vectors that appear in the original module. For both $j=j_0\pm k$,
  the module generated by $j=j_0$ is not a submodule (i.e. $j_0$ not a
  singular vector within the original module). This can be explicitly checked
  by finding solutions to the 
Kac-Kazhdan equation in \eqref{kackazhdan}.
\end{itemize}
This feature generalizes to the generic case with more nodes and also to the
rational case, and has inspired our ansatz in \eqref{pbyqansatz} which we have proven in the main text using the method of induction. 
\end{appendix}

\bibliographystyle{unsrt}

\begin{thebibliography}{10}

\bibitem{Kiritsis:1993pb}
E.~Kiritsis, C.~Kounnas and D.~Lust,
``A Large class of new gravitational and axionic backgrounds for
four-dimensional superstrings,''
Int.\ J.\ Mod.\ Phys.\ A {\bf 9} (1994) 1361
[arXiv:hep-th/9308124].

\bibitem{Hori:2002cd}
K.~Hori and A.~Kapustin,
``Worldsheet descriptions of wrapped NS five-branes,''
JHEP {\bf 0211}, 038 (2002)
[arXiv:hep-th/0203147].

\bibitem{Giveon:1999px}
A.~Giveon and D.~Kutasov,
``Little string theory in a double scaling limit,''
JHEP {\bf 9910}, 034 (1999)
[arXiv:hep-th/9909110].

\bibitem{Hori:2001ax}
K.~Hori and A.~Kapustin,
``Duality of the fermionic 2d black hole and N = 2 Liouville theory as  mirror
symmetry,''
JHEP {\bf 0108} (2001) 045
[arXiv:hep-th/0104202].

\bibitem{Tong:2003ik}
D.~Tong,
``Mirror mirror on the wall: On two-dimensional black holes and Liouville
theory,''
JHEP {\bf 0304}, 031 (2003)
[arXiv:hep-th/0303151].

\bibitem{Israel:2004jt}
D.~Israel, A.~Pakman and J.~Troost,
``D-branes in N = 2 Liouville theory and its mirror,''
Nucl.\ Phys.\ B {\bf 710}, 529 (2005)
[arXiv:hep-th/0405259].

\bibitem{kutasov}
  D.~Kutasov and N.~Seiberg,
  ``Noncritical Superstrings,''
Phys.\ Lett.\ B {\bf 251}, 67 (1990).

\bibitem{Mizoguchi:2000kk}
S.~Mizoguchi,
``Modular invariant critical superstrings on four-dimensional Minkowski  space
x two-dimensional black hole,''
JHEP {\bf 0004}, 014 (2000)
[arXiv:hep-th/0003053].

\bibitem{Eguchi:2000tc}
T.~Eguchi and Y.~Sugawara,
``Modular invariance in superstring on Calabi-Yau n-fold with A-D-E
singularity,''
Nucl.\ Phys.\ B {\bf 577}, 3 (2000)
[arXiv:hep-th/0002100].

\bibitem{Murthy:2003es}
S.~Murthy,
``Notes on non-critical superstrings in various dimensions,''
JHEP {\bf 0311}, 056 (2003)
[arXiv:hep-th/0305197].

\bibitem{Eguchi:2004yi}
T.~Eguchi and Y.~Sugawara,
``SL(2,R)/U(1) supercoset and elliptic genera of non-compact Calabi-Yau
manifolds,''
JHEP {\bf 0405}, 014 (2004)
[arXiv:hep-th/0403193].

\bibitem{Israel:2004ir}
D.~Israel, C.~Kounnas, A.~Pakman and J.~Troost,
``The partition function of the supersymmetric two-dimensional black hole  and
little string theory,''
JHEP {\bf 0406}, 033 (2004)
[arXiv:hep-th/0403237].

\bibitem{Eguchi:2004ik}
T.~Eguchi and Y.~Sugawara,
``Conifold type singularities, N = 2 Liouville and SL(2,R)/U(1) theories,''
JHEP {\bf 0501}, 027 (2005)
[arXiv:hep-th/0411041].

\bibitem{giveon2}
  A.~Giveon, D.~Kutasov and O.~Pelc,
  ``Holography for non-critical superstrings,''
  JHEP {\bf 9910}, 035 (1999) [arXiv:hep-th/9907178].

\bibitem{Mukhi}
  S.~Mukhi and C.~Vafa,
 ``Two-dimensional black hole as a topological coset model of c = 1 string theory,''  Nucl.\ Phys.\ B {\bf 407}, 667 (1993)  [arXiv:hep-th/9301083].

\bibitem{Ghoshal}
  D.~Ghoshal and C.~Vafa,
  ``C = 1 string as the topological theory of the conifold,''
  Nucl.\ Phys.\ B {\bf 453}, 121 (1995)
  [arXiv:hep-th/9506122].

\bibitem{bertoldi}
  G.~Bertoldi, 
  ``Double scaling limits and twisted non-critical superstrings,'' [arXiv:hep-th/0603075].

\bibitem{Gato-Rivera:1992fd}
B.~Gato-Rivera and A.~M.~Semikhatov,
``$d \le 1\cup d \ge 25$ and W constraints from BRST invariance in the $c \ne 3$
topological algebra,''
Phys.\ Lett.\ B {\bf 293} (1992) 72
[Theor.\ Math.\ Phys.\  {\bf 95} (1993\ TMPHA,95,535-545.1993\ TMFZA,95,239-250.1993) 536]
[arXiv:hep-th/9207004].

\bibitem{Semikhatov}
  A.~M.~Semikhatov,
 ``The MFF singular vectors in topological conformal theories,''
  Mod.\ Phys.\ Lett.\ A {\bf 9}, 1867 (1994)  [JETP Lett.\  {\bf 58}, 860 (1993)] [arXiv:hep-th/9311180].

\bibitem{Feigin:1997ha}
B.~L.~Feigin, A.~M.~Semikhatov and I.~Y.~Tipunin,
``Equivalence between chain categories of representations of affine sl(2)  and
N = 2 superconformal algebras,''
J.\ Math.\ Phys.\  {\bf 39}, 3865 (1998)
[arXiv:hep-th/9701043].

\bibitem{Takayanagi}
  T.~Takayanagi,
  ``$c < 1$ string from two dimensional black holes,''
  arXiv:hep-th/0503237.

\bibitem{Sahakyan}
  D.~A.~Sahakyan and T.~Takayanagi,
  ``On the connection between N = 2 minimal string and (1,n) bosonic minimal string,'' [arXiv:hep-th/0512112].

\bibitem{Niarchos}
  V.~Niarchos,
  ``On minimal N = 4 topological strings and the (1,k) minimal bosonic string,''
  arXiv:hep-th/0512222.

\bibitem{Rastelli} 
L.~Rastelli and M.~Wijnholt, ``Minimal AdS(3),'' arXiv:hep-th/0507037.

\bibitem{Kazama}
  Y.~Kazama and H.~Suzuki,
  ``New N=2 Superconformal Field Theories And Superstring Compactification,''   Nucl.\ Phys.\ B {\bf 321}, 232 (1989).

\bibitem{Witten}
  E.~Witten,
  ``On string theory and black holes,''
  Phys.\ Rev.\ D {\bf 44}, 314 (1991).

\bibitem{Dijkgraaf}
  R.~Dijkgraaf, H.~L.~Verlinde and E.~P.~Verlinde,
  ``String propagation in a black hole geometry,''
  Nucl.\ Phys.\ B {\bf 371}, 269 (1992).

\bibitem{Figueroa-O'Farrill:1995pv}
J.~M.~Figueroa-O'Farrill and S.~Stanciu,
``N=1 and N=2 cosets from gauged supersymmetric WZW models,''
arXiv:hep-th/9511229.
\bibitem{Karabali:1988au}
D.~Karabali, Q.~H.~Park, H.~J.~Schnitzer and Z.~Yang,
Phys.\ Lett.\ B {\bf 216} (1989) 307.

\bibitem{Hwang:1993nc}
S.~Hwang and H.~Rhedin,
``The BRST Formulation of G/H WZNW models,''
Nucl.\ Phys.\ B {\bf 406} (1993) 165
[arXiv:hep-th/9305174].

\bibitem{Bouwknegt}
  P.~Bouwknegt, J.~G.~McCarthy and K.~Pilch,
  ``BRST analysis of physical states for 2-D gravity coupled to $c \le 1$ matter,''
  Commun.\ Math.\ Phys.\  {\bf 145}, 541 (1992).

\bibitem{Ohta}
  N.~Ohta and H.~Suzuki,
 ``Bosonization of a topological coset model and noncritical string theory,''
  Mod.\ Phys.\ Lett.\ A {\bf 9}, 541 (1994) [arXiv:hep-th/9310180].

\bibitem{Frenkel:1992ex}
E.~Frenkel,
``Determinant formulas for the free field representations of the Virasoro and
Kac-Moody algebras,''
Phys.\ Lett.\ B {\bf 286} (1992) 71.

\bibitem{Malikov}
F. Malikov, Algebra y Analiz, 2 No. 2, 65 (1990).

\bibitem{Feigin:1990ut}
B.~L.~Feigin and E.~V.~Frenkel,
``Representations Of Affine Kac-Moody Algebras And Bosonization,''

\bibitem{Kac}
  V.~G.~Kac and D.~A.~Kazhdan,
  ``Structure Of Representations With Highest Weight Of Infinite Dimensional Lie Algebras,''
  Adv.\ Math.\  {\bf 34}, 97 (1979).

\bibitem{Ashok:2005xc}
  S.~K.~Ashok, S.~Murthy and J.~Troost,
  ``Topological cigar and the c = 1 string: Open and closed,'' JHEP {\bf 0602}, 013 (2006)
  [arXiv:hep-th/0511239].

\bibitem{MFF}
F.~G.~Malikov, B.~L.~ Feigin and D.~B.~Fuchs,
`` Singular vectors in Verma Modules over Kac-Moody algebras'',
Funk. An. Prilozh. {\bf 20} No 2, 25 (1986).

\bibitem{Bauer:1993jj}
M.~Bauer and N.~Sochen,
Commun.\ Math.\ Phys.\  {\bf 152} (1993) 127
[arXiv:hep-th/9201079].

\bibitem{Bershadsky:1993cx}
M.~Bershadsky, S.~Cecotti, H.~Ooguri and C.~Vafa,
``Kodaira-Spencer theory of gravity and exact results for quantum string
amplitudes,''
Commun.\ Math.\ Phys.\  {\bf 165} (1994) 311
[arXiv:hep-th/9309140].

\bibitem{Lerche:1989uy}
W.~Lerche, C.~Vafa and N.~P.~Warner,
``Chiral Rings In N=2 Superconformal Theories,''
Nucl.\ Phys.\ B {\bf 324}, 427 (1989).

\bibitem{Aganagic}
  M.~Aganagic, R.~Dijkgraaf, A.~Klemm, M.~Marino and C.~Vafa,
  ``Topological strings and integrable hierarchies,''
  Commun.\ Math.\ Phys.\  {\bf 261}, 451 (2006)  [arXiv:hep-th/0312085].

\bibitem{DiFrancesco:1997nk}
P.~Di Francesco, P.~Mathieu and D.~Senechal,
``Conformal field theory,'', Springer-Verlag New-York, 1997.


\end{thebibliography}

\end{document}